\documentclass{vgtc}                          

\ifpdf
  \pdfoutput=1\relax                   
  \usepackage{graphicx}                
  \DeclareGraphicsExtensions{.pdf,.png,.jpg,.jpeg} 
\else
  \usepackage{graphicx}                
  \DeclareGraphicsExtensions{.eps}     
\fi%

\graphicspath{{figures/}{pictures/}{images/}{./}} 

\usepackage{microtype}                 
\PassOptionsToPackage{warn}{textcomp}  
\usepackage{textcomp}                  
\usepackage{mathptmx}                  
\usepackage{times}                     
\usepackage{cite}                      
\usepackage{tabu}                      
\usepackage{booktabs}                  
\usepackage{soul}
\usepackage{todonotes}
\usepackage{balance}

\sethlcolor{cyan}
\renewcommand\hl[1]{#1}

\onlineid{1285}

\vgtccategory{Research}

\vgtcinsertpkg

\title{Pen-based Interaction with Spreadsheets in Mobile Virtual Reality}

\author{
 Travis Gesslein$^{1}$
 \and Verena Biener$^{1}$
\and Philipp Gagel$^{1}$
\and Daniel Schneider$^{1}$ 
\and Per Ola Kristensson$^{3}$
\and Eyal Ofek$^{2}$
\and Michel Pahud$^{2}$
\and Jens Grubert$^{1}$\thanks{contact author: jens.grubert@hs-coburg.de}
}
\vspace{-0.15cm}
\affiliation{\scriptsize $^{1}$Coburg University of Applied Sciences and Arts  $^{2}$Microsoft Research  \\ $^{3}$University of Cambridge}

\teaser{
\centering
 \includegraphics[width=0.8\columnwidth]{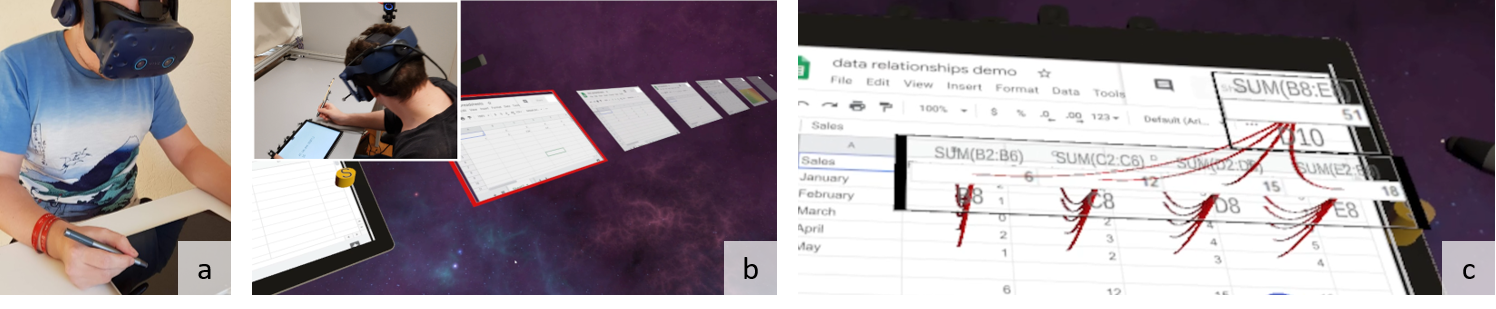}
 \caption{This paper investigates the joint interaction space between an immersive virtual reality headset, and a spatially tracked pen for interacting with spreadsheets on a mobile tablet (a). This approach allows spreadsheet data to be extended around the screen and, for example, enables the user to interact with adjacent spreadsheets (b). It also allows additional information to be provided above the screen and, for example, enables the system to unravel hidden dependencies between cells by directly visualizing the cell relationships in 3D space (c).  }
 \label{fig:teaser}
}

\abstract{Virtual Reality (VR) can enhance the display and interaction of mobile knowledge work and in particular,  spreadsheet applications. 
While spreadsheets are widely used yet are challenging to interact with, especially on mobile devices, using them in VR has not been explored in depth.
A special uniqueness of the domain is the contrast between the immersive and large display space afforded by VR, contrasted by the very limited interaction space that may be afforded for the information worker on the go, such as an airplane seat or a small work-space. To close this gap, we present a tool-set for enhancing spreadsheet interaction on tablets using immersive VR headsets and pen-based input. This combination opens up many possibilities for enhancing the productivity for spreadsheet interaction. 
We propose to use the space around and in front of the tablet for enhanced visualization of spreadsheet data and meta-data. For example, extending sheet display beyond the bounds of the physical screen, or easier debugging by uncovering hidden dependencies between sheet's cells. Combining the precise on-screen input of a pen with spatial sensing around the tablet, we propose tools for the efficient creation and editing of spreadsheets functions such as off-the-screen layered menus, visualization of sheets dependencies, and \hl{gaze-and-touch-based} switching between spreadsheet tabs.
We study the feasibility of the proposed tool-set using a video-based online survey and an expert-based assessment of indicative human performance potential.

} 

\begin{document}


\firstsection{Introduction}

\maketitle

Spreadsheets are widely used data modeling, manipulation, and storage tools that are used in many application areas \cite{birch2018future}. Interestingly, the basics of spreadsheet interaction have largely remained unchanged over the last 30 years. 
Spreadsheet popularity stems from their ease-to-learn, simple principles, and their flexibility \cite{birch2018future}. Yet these characteristics are also the spreadsheets' limitations.

Unlike many other applications, spreadsheets are inherently open, and, due to their uniform grid nature, simple to learn for beginners. For example, navigating the sheet and copying cells are the same actions regardless of the sheet's content. However, the same uniformity also makes it hard to identify the structure of the spreadsheet or to debug complex dependencies in the spreadsheet since it does not allow a display of any type of information that does not conform to the grid. Previous research efforts have tried to visualize the hierarchy of the connection structure using 3D renderings \cite{Shiozawa1999, Hermans2011}, yet any stray from the regular grid display---the user's main workplace---has not managed to achieve wide user adoption.

The spreadsheet also acts as an unconstrained canvas for bringing diverse ideas to life. It might become a simple to-do list, a calendar, a financial model, or a 500,000 cell encompassing inter-disciplinary project. The unconstrained canvas means that no matter how big the project, there is enough sheet space to accommodate it.  This flexibility, while it is the spreadsheet’s biggest asset, frequently proves to be it’s Achilles’ heel as well. Users often find themselves in need to scroll again and again along with spreadsheets much larger than the size of their screens. Large chunks of data may be needed to be selected, and, to be assigned as an input to functions multiple times. A possible mitigation strategy is using a larger screen that can show more data as the size of a mobile display proves to be too small. An overview of the spreadsheet may also help to access data outside the screen. However, currently, all available screen space is used for displaying as much of the spreadsheet as possible.   

Further, this interaction paradigm---selecting cells, entering data by symbolic input, navigating around an unconstrained canvas---has been carried over from traditional desktop-based environments, which use a physical keyboard and mouse as standard input devices, to mobile settings, where users operate smaller-sized touchscreen devices, such as tablets or smartphones via touch and pen-based input \cite{flood2011useful}. Both the small screen space and the data entry methods using touch lead to a number of usability problems, first and foremost increased error rates when interacting with spreadsheet data \cite{flood2011spreadsheets}. Still, there is a strong need for accessing spreadsheet data when away from traditional desktop-based computing devices \cite{flood2011spreadsheets}.

We see potential in facilitating interaction with spreadsheets in mobile settings using Virtual Reality (VR). First, VR head-mounted displays (HMDs) can help users visualize a much larger display area around them. As a handheld tablet may have the angular field of view of approximately 30 degrees, the accumulative horizontal field of view of VR HMDs can span 100-180 degrees and more. While users may still use the familiar regular grid as their main tool of interaction, this additional display volume, in contrast to the mobile's small and flat screen, can also reveal information outside the screen. It allows the system to visualize additional data the user can interact with, around and above the main spreadsheet, see Figure \ref{fig:teaser}. Underpinned by the spatial sensing capabilities of modern pen-operated tablets and hover sensing of the pen \cite{Fitzmaurice03trackingmenus, grossmann2006hover}, we set out to explore the joint interaction space of VR HMDs and pen-based inputs for interacting with spreadsheets on mobile devices, such as tablets.

In this work, we are not trying to inherently change the nature of spreadsheets. Instead, our objective is to show that embedding spreadsheet interaction within a 3D space can help in exposing the internal structure and allow better interaction possibilities. While there are many aspects and applications that can be analyzed, we chose a number of selected techniques that demonstrate the advantages of using VR for the mobile knowledge worker. While we extend the display of information in front, behind, and around the tablet screen, we do keep the interaction mostly in the limited area of the tablet. The 2D nature of the input device is complemented by pointing and editing information on or near the screen surface by hands that are rested on the tablet or an underlying table, simply by tilting the wrist to raise the pen. 

 
In this paper, we make two contributions: 1) we design and implement techniques that combine VR and pen-based input on and above tablets for efficient interaction with spreadsheets, and 2) we validate those techniques in two indicative studies.

\section{Related Work}

\begin{figure}[t!]
  \centering
  \includegraphics[width=\columnwidth]{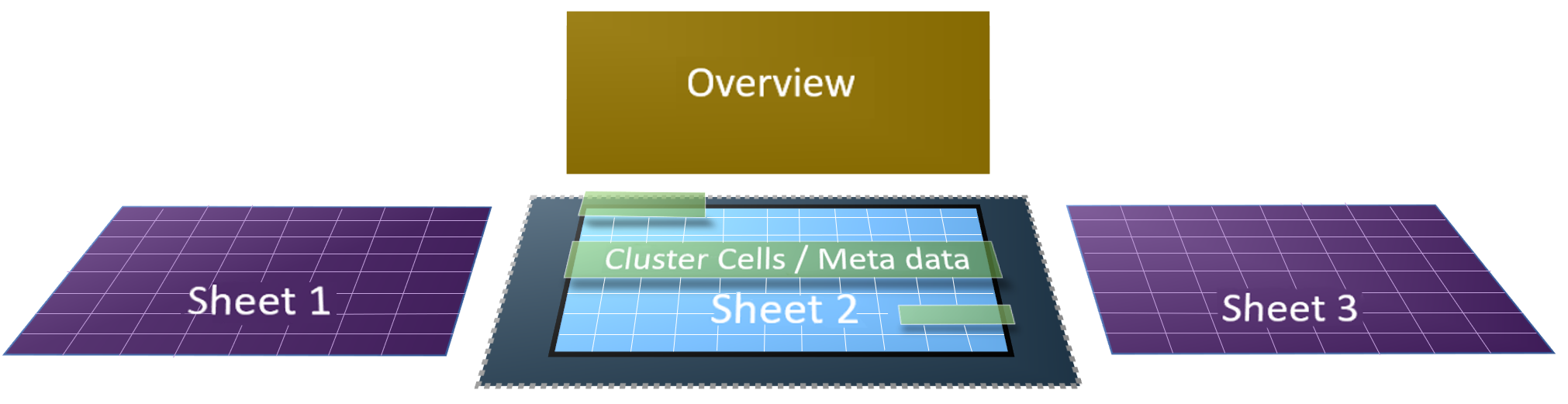}
  \caption{Concept for utilizing the space around the main spreadsheet (light blue) which spatially coincides with the tablet: The sheet area can be extended (dark blue, dotted lines) beyond the tablet bounds. Additional sheets can be visualized on both sides of the active sheet (purple). Additional data layers can be placed at corresponding (x,y)-positions facilitating faster data association. Behind the sheet additional windows (yellow) can show an overview visualization.}~\label{fig:spatialarrangement}.
  \vspace{-0.8cm}
\end{figure}

Our work draws upon the rich resources in spreadsheet interaction, pen-based and in-air interaction, context-menus and VR for knowledge workers.  

\subsection{Spreadsheet Interaction}
According to Burnett et al.~\cite{burnett2001forms}, the spreadsheet is probably the most popular programming paradigm in use, even though it presents several limitations and challenges to users. Prior work has identified challenges when interacting with spreadsheet software. For example, Mack et al.~\cite{mack2018characterizing} collected complaints from Reddit and characterized issues users were facing. They identified challenges about tasks, such as importing, managing, querying, and presenting data. Smith et al.~\cite{smith2017spreadsheet} investigated both individual and organizational challenges in using spreadsheet software, and, amongst others, identified data pipeline challenges related to importing data. 
Chambers et al.~\cite{chambers2010struggling} describes challenges identified by spreadsheet users in a field study and proposed new functions including developing different modes for spreadsheet creation, improving support for spreadsheet reuse, and helping users to find and use features. Flood et al.~\cite{flood2008evaluation} identified navigation as an issue that affects the performance of users debugging spreadsheets. Birch et al.~\cite{birch2018future} described success factors of current spreadsheet technologies but also challenges such as hidden errors, comprehensibility, and complexity. Specifically, regarding comprehensibility Birch et al.~\cite{birch2018future} note that ``One underlying reason for this high error rate is known to be the user’s difficulty in understanding the spreadsheet they are interacting with \cite{panko2016we, kohlhase2015context}'', that is, that formulas are hidden by default and that the wider hidden structure of a spreadsheet model is not visible, even if formulas (and their first-order dependent cells) are shown.

Further, several researchers have investigated novel interaction methods and models for spreadsheet use. For example, Miller et al.~\cite{miller2016gradual} demonstrate novel features that enable the gradual structuring of spreadsheets based on design patterns of expert spreadsheet modelers. Jones et al.~\cite{jones2003user} applied Cognitive Dimensions of Notations \cite{blackwell2003notational} to their proposal for adding user-defined functions, also called sheet-defined functions, to spreadsheets. Janach et al.~\cite{jannach2016model} adopt techniques from model-based diagnosis to spreadsheet debugging. Kandogan et al.~\cite{kandogan2005a1} present a spreadsheet-based environment with a task-specific system-administration language for quickly creating small tools or migrating existing scripts to run as web portlets.

Regarding the use of mobile spreadsheet applications, Flood et al.~\cite{flood2011spreadsheets} identified a strong need for accessing spreadsheet data when away from traditional desktop-based computing devices. Chintapalli et al.~\cite{chintapalli2016comparative} compared four mobile spreadsheet applications according to the following usability criteria: visibility, navigation, scrolling, feedback, interaction, satisfaction, simplicity, and convenience. They found few differences among those applications but identified visibility, navigation, and feedback challenges. In contrast, a systematic review of mobile spreadsheet applications~\cite{flood2011systematic} revealed substantial differences in available functions (such as the ability to sort data or to keep headings visible while scrolling) between mobile spreadsheet applications. Flood et al.~\cite{flood2011useful} also identified further challenges mobile users faced when using spreadsheet applications on smartphones, such as inaccurate cell and character selection, unintended actions, and unexpected behaviors. 

Based on the insights for both the need for interacting with spreadsheets in mobile settings but also its challenges due to limited input and output capabilities~\cite{flood2011spreadsheets}, we investigate how the joint interaction space between HMDs and pen-based input can support interaction with spreadsheets on tablets.

\subsection{Mixed Reality for Knowledge Worker Tasks}

The use of Mixed Reality (MR) for supporting knowledge work has attracted recent research interest \cite{grubert2018office, ruvimova2020transport, guo2019mixed}. While early work investigated projection systems to extend physical office environments  (e.g., \cite{wellner1994interacting, kobayashi1998enhanceddesk, rekimoto1999augmented, pinhanez2001everywhere}), more recently, VR and AR HMDs have been investigated as tools for assisting users with interacting with physical documents (e.g., \cite{grasset2007mixed, li2019holodoc}), focusing on annotating documents displayed on a 2D surfaces.
Grubert et al. \cite{grubert2018office} and McGill et al. \cite{mcgill2019challenges} explored the positive and negative qualities that VR introduces in mobile scenarios on the go. Other works has investigated VR use in desktop-based environments for tasks such as text entry (e.g., \cite{mcgill2015dose, knierim2018physical, grubert2018text}), system control \cite{zielasko2019passive, zielasko2019menus} and visual analytics \cite{wagner2018virtualdesk, buschel2018interaction}. Research on productivity-oriented desktop-based VR has concentrated on the use of physical keyboards \cite{schneider2019reconviguration}, controllers and hands \cite{kry2008handnavigator, zielasko2019passive}, and, recently, tablets \cite{surale2019tabletinvr}. Concurrently with our work, Biener et al. \cite{Biener2020Breaking} investigated the joint interaction space of VR HMDs and tablets for a variety of knowledge worker tasks.

We complement this prior work by investigating a specific information worker application -- spreadsheets using tablets -- in detail.

\begin{figure*}[t!]
  \centering
  \includegraphics[width=2\columnwidth]{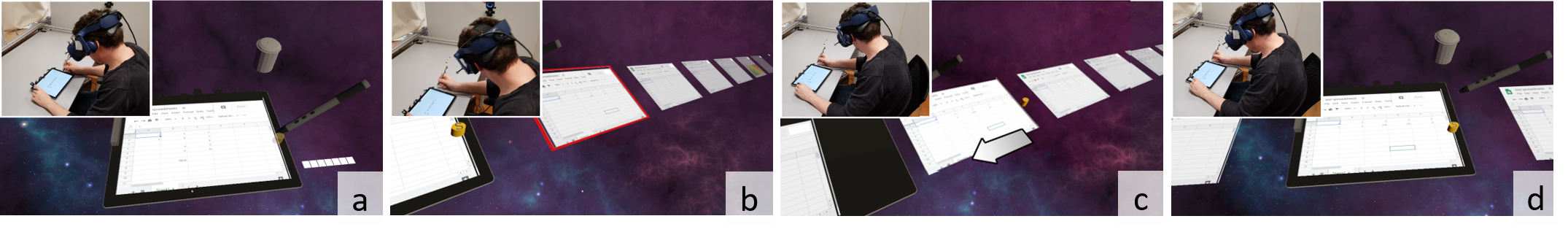}
  \caption{Interacting with multiple sheets using: initially, solely icons indicating the existent of additionally accessible sheets are visible (a). Neighboring sheets are expanded and each sheet the user gazes at is highlighted with a red frame (b). The user taps with his non-dominant hand on the tablet bezel, causing the selected sheet to slide towards the tablet (c), where the user can edit it using the tablet's touchscreen.}~\label{fig:multisheet-eye}
  \vspace{-0.7cm}
\end{figure*}

\begin{figure}[t!]
  \centering
  \includegraphics[width=0.8\columnwidth]{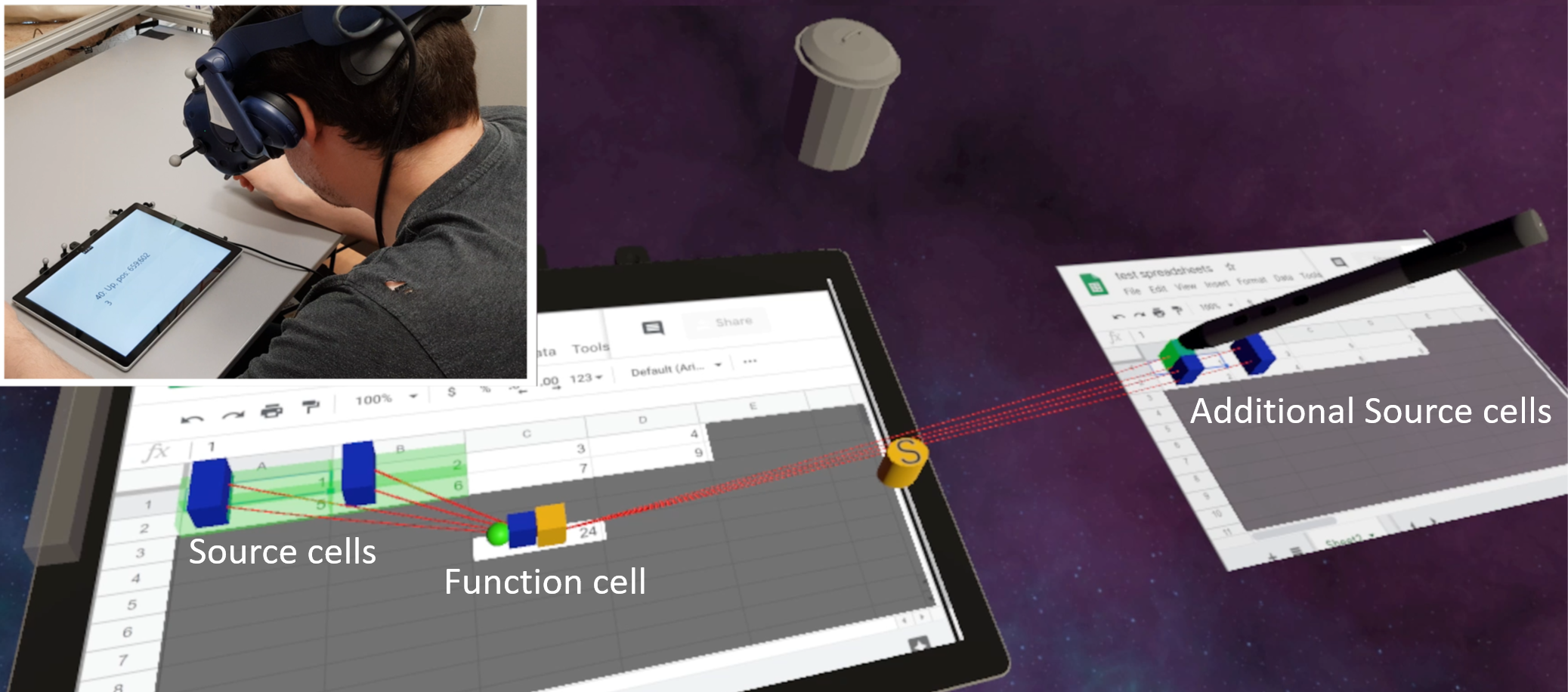}
  \caption{Neighboring sheets can also be accessed via in-air pointing. Here, the user extends the contribution to a function he defines, adding a cell of a neighboring sheet.}~\label{fig:multisheet-ia}
  \vspace{-0.9cm}
\end{figure}

\subsection{Pen-based and In-Air Interaction}


Besides the commonly used single-point input with pens, enhanced interaction techniques have been explored. Examples include using touch input on the non-dominant hand, supporting pen input in bimanual interaction (e.g.,~\cite{brandl2008combining, hinckley2010pen, matulic2013pen, pfeuffer2017thumb}, unimodal surface-based pen-postures~\cite{cami2018unimanual}, bending~\cite{fellion2017flexstylus} or using sensors in or around the pen~\cite{hasan2012acoord, liu2012flexaura, hinckley2013motion, hwang2013magpen, teyssier2017versapen, matulic2020pensight} for gestures and postures, and examining pen-grips (e.g.,~\cite{suzuki2009interaction, hyong2011grips, hinckley2014sensing}). Our work was inspired by tilting~\cite{tian2008tilt} and hovering~\cite{Fitzmaurice03trackingmenus, grossmann2006hover} the pen above interactive surfaces, which we use in a VR context. 


The use of pens in AR and VR has also been investigated as a standard input device on physical props \cite{szalavari1997personal, szalavari1997using}, as well as using grip-specific gestures for mid-air interaction \cite{li2020grip}. The accuracy of pen-based mid-air pointing has also been studied \cite{pham2019pen, batmaz2020precision}.

Regarding prior work on combining in-air with touch interaction, Marquardt et al.~\cite{marquardt2011continuous} investigated the use of on and above surface input on a tabletop. Chen et al.~\cite{chen2014air+} explored in-air use of on and above surface input on a tabletop. They propose that interactions can be composed by interweaving in-air gestures before, between, and after touch on a prototype smartphone augmented with hover sensing. Hilliges et al.~\cite{hilliges2009interactions} have been using hover to allow more intuitive interaction with virtual objects that represent physical objects. More recently, Hinckley et al.~\cite{hinckley2016pre-touch} have been exploring a pre-touch modality on a smartphone including the approach trajectories of fingers to distinguish between different operations. Such technology can be used to connect 3D tracking and touchscreen digitizer for better accuracy of tracking.

Most VR in-air interaction typically aims at using unsupported hands. To enable reliable selection, targets are designed to be sufficiently large and spaced apart \cite{speicher2018VRselection}. Our focus on mobile knowledge workers on the move dictates small gestures to reduce working fatigue and to retain operationalizability in potentially cramped environments, such as airplane seats. 
We design gestures to be used by a hand, resting on the screen of a tablet and holding a pen. 
The pen, or stylus, is a tool that is designed for writing, but also allow precise operation and selection \cite{hinckley2007inkseine} and has buttons to trigger actions, in addition to enabling using handwriting recognition in cells in future extensions of this work.
Pen gestures use the fine finger motions to enable fine selection among selections on the screen-surface or above it, as the pen can be tilted up and down to select between layers of menus in the vertical direction to enable new interactions. 
For example, when using in-place 2D menus, e.g. pie menus, where each selection opens another sub-menu, returning to a parent menu may require a designated gesture. In contrast, in a 3D space above the tablet screen, the user can simply tilt the pen toward a lower menu to re-select it.

\subsection{Context Menus, In-place Commands and Bimanual Interaction} 

In-place or at-hand commands appearing next to the user on demand have the benefits 
of avoiding a trip to a fixed menu on the screen. Bier et al.~\cite{bier1993} explored the benefits of toolglasses and magic lenses manipulated indirectly using a mouse on the preferred-hand, and a trackball/thumb-wheel on the non-preferred-hand. In-place commands also work with direct manipulations on modern pen-and-touch displays by placing the menu near non-preferred-hand fingers and using the pen in the preferred-hand to consume the menu. These types of menus can be very useful on large displays where menus may be out of reach. Xia et al.~\cite{xia2017writlarge} used an 84 inches Microsoft Surface Hub to explore, seamlessly select or frame content with the non-preferred-hand and manipulate it with the preferred-hand using marking menus \cite{Kurtenbach94userlearning} appearing in-context at the border of the selection area. Further, large displays' in-place commands can allow multiple users to work together side-by-side \cite{webb2016wearables}. On-demand in-place commands are also useful for small displays, where a screen might have insufficient real-estate to display a fixed menu. A thumb menu may be a special on-demand menu for a hand holding a tablet or a smartphone \cite{pfeuffer2017thumb}. Different menu types, such as floating palettes, marking menus, and toolglasses have been investigated for their strengths and weaknesses \cite{mackay2002interactionTechnique} and lately have been explored in VR \cite{lim2019evaluation}. In addition, Tian et al.~\cite{tian2008tilt} have been looking into using the orientation (tilt) of a pen to select menus while it's tip is positioned statically on the screen.

Our work complements such prior work by using 3D layered hierarchical pie-menu in close vicinity to a resting touch surface. It allows small hand gestures to easily select among multiple menus.


\begin{figure*}[t!]
  \centering
  \includegraphics[width=2\columnwidth]{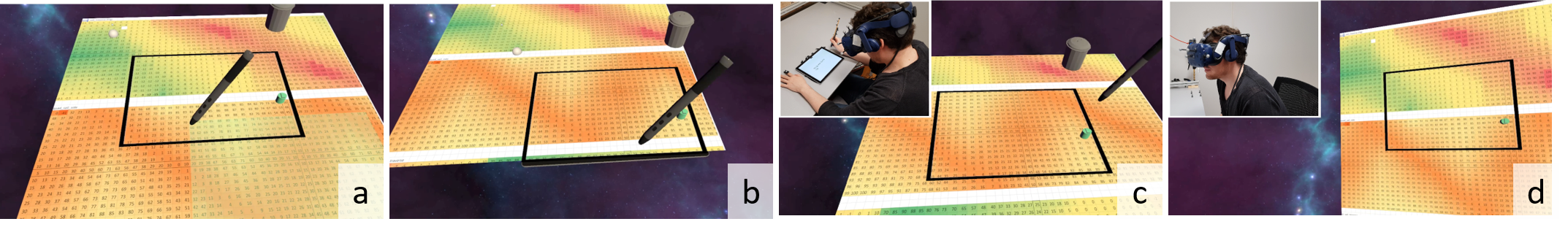}
  \caption{Extended view of a single sheet beyond the screen area (the semi-transparent green rectangular viewport) (a) By selecting a cell outside the viewport, the spreadsheet data may slide to be aligned with the touchscreen (b). Another option is to keep the viewport fixed in order to be able to reach to all sides of the touchscreen (c). Alternatively, the coordinate system of the interactive space can be tilted to be vertical, allowing a more comfortable display of a large sheet area in front of the user, while her physical hands are moving horizontally over the tablet (d).}~\label{fig:extended-view}
\end{figure*}

\begin{figure}[t!]
  \centering
  \includegraphics[width=0.8\columnwidth]{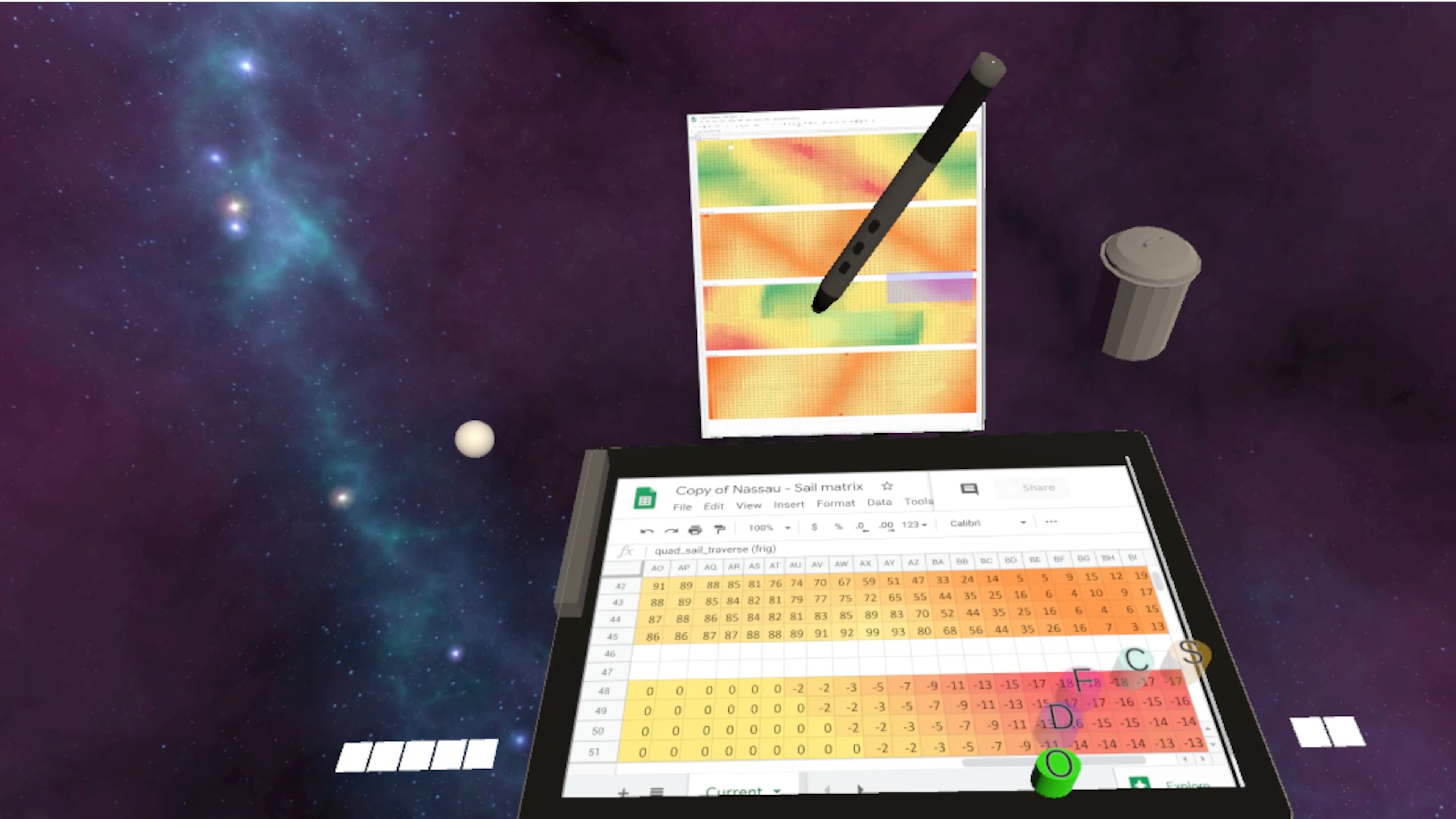}
  \caption{Overview visualization of the current sheet placed behind the active sheet.}~\label{fig:overview-vis}
\end{figure}

\section{Spreadsheet Interaction}

Spreadsheet interaction has been mostly unchanged for decades: scrolling around a 2D grid using mouse or keyboard commands, interaction with cells via the keyboard, mostly on an edit-line outside the grid, and selecting and copying cells using the mouse. 

In contrast, working with a pen on a tablet \hl{or on a horizontal interactive surface} is an interaction style of a different nature \hl{(e.g.,} \cite{brandl2008combining, hinckley2010pen, matulic2013pen, pfeuffer2017thumb}\hl{)}.
To finely control the pen, the palm rests on the tablet screen. Moving the hand around the screen incurs a higher cost than sliding a mouse, resulting in a high incentive to bring the interaction \hl{and menus} to the vicinity of the pen's reach \hl{(e.g.,} \cite{zhang2019sensing}\hl{)}. 
Using HMDs allows for positioning additional display real-estate near the hand without interfering with the 2D grid content by exploiting the 3D space around the hand.  



In this work, we propose a set of tools for (1) enhancing the visibility of spreadsheet elements and meta data around the physical screen using VR; and (2) streamlining workflows for creating and editing spreadsheet functions using pen-based, multimodal interaction techniques. 
We designed the techniques using an iterative approach with multiple design iterations consisting of conceptualization, implementation and informal user tests ('eat your own dogfood' internal testing) \cite{unger2012project, drachen2018games}.

\subsection{Around Tablet Visualization}


Our concept uses the 3D space, around, above, and behind the tablet, to display additional information, both extended views of the spreadsheet as well as in-place contextual information (Figure \ref{fig:spatialarrangement}). While the area of the spreadsheet used for 2D interaction lies mostly on the tablet screen area, the extended accumulated field-of-view of the HMD enables extending the visible area of the grid (dark blue area in Figure \ref{fig:spatialarrangement}), as well as displaying additional tabs (purple sheets floating next to the tablet in Figure \ref{fig:spatialarrangement}). Such tabs can easily become replaced to be the edited tab on the tablet screen using a selection technique that combines eye-gaze and touch gestures. The space beyond the tablet can display additional information, such as a zoom-out overview of the spreadsheet, thereby enabling better discoverability of dependent elements outside the current field of view, and in addition provide fast navigation. Finally, the 3D space above the tablet's screen area is used to display multiple layers of contextual information, in-place, relative to the 2D spreadsheet.

\subsubsection{Interaction with Multiple Tabs}

Widely used spreadsheet software enables an arrangement of a large spreadsheet into separate tabs, accessible at the bottom of the spreadsheet window. However, as only one tab may be visible at a time, it may hamper visual reference and linking of data across tabs. Using the large field of view of immersive VR HMDs, we propose to display multiple tabs, extending them to the side of the tablet (Figure \ref{fig:multisheet-eye}, b), allowing for an easy association of data between neighboring tabs. 

As only the current tab aligns with the physical tablet screen, we support two techniques for interacting with tabs. A combined gaze-and-touch interaction, \hl{using head-gaze}, lets the user look at any tab, \hl{which is highlighted when the ray representing the head direction hits this tab, }(the red frame shown in Figure \ref{fig:multisheet-eye}, b) and tapping a non-dominant-hand finger on the touchscreen bezel will slide the tabs until the selected tab is aligned with the touchscreen  (Figure \ref{fig:multisheet-eye}, c, and d). Alternatively, neighboring sheets can directly be accessed by \textit{in-air} pointing with the pen (Figure \ref{fig:multisheet-ia}).

Both techniques can be seamlessly combined. While gaze-and-touch enables access of out-of-reach tabs and support of the tablet when editing, it requires time for tab selection. Furthermore, frequent switching of sheets may increase simulator-sickness risk due to frequent head rotations. In-air interaction provides faster access to neighboring tabs within the user's arm's reach, but has less input fidelity due to the lack of screen support and sensing (which can be somewhat mitigated using a table as a resting surface.)


\subsubsection{Extending a Single Sheet}

In typical spreadsheet applications, accessing cells outside of the view-area of the display's viewport requires scrolling the sheet to reveal new unseen areas. A VR extended display space enables us to show a larger virtual screen (Figure \ref{fig:extended-view}), however, only part of the display is supported by the tablet's screen sensing. To enable editing of the entire sheet, it is possible to select a different region of the sheet using gaze-and-touch to slide the region until it is aligned with the tablet area. Another option is to re-target the viewport, as well as the view of the virtual pen from their location aligned with the physical screen toward the area to be edited \cite{grubert2018text}. Visually, it appears to the user as if the tablet's screen moves from its original location. When using a very large displayed spreadsheet, it may be better to re-target it into a vertical display to avoid the sensation of the plane penetrating the user's body.  

A complementary approach may use an overview visualization of the spreadsheet placed behind the tablet (Figure \ref{fig:overview-vis}). Users can select a region of interest with their pen, sliding the sheet to align the selected location with the physical touchscreen.


\subsection{On and Above Tablet Visualization}

Enabling the display of additional information beyond the regular grid format can help to show the user important information, such as the flow of data along with the sheet, intermediate debugging results, and other meta information. We propose to use the space above the tablet surface, which is still reachable by the user's hand tilting the pen, to display different semantic information in proximity to the grid cell's content.

\subsubsection{Visualizing Hidden Dependencies}

\begin{figure*}[tb!]
  \centering
  \includegraphics[width=2\columnwidth]{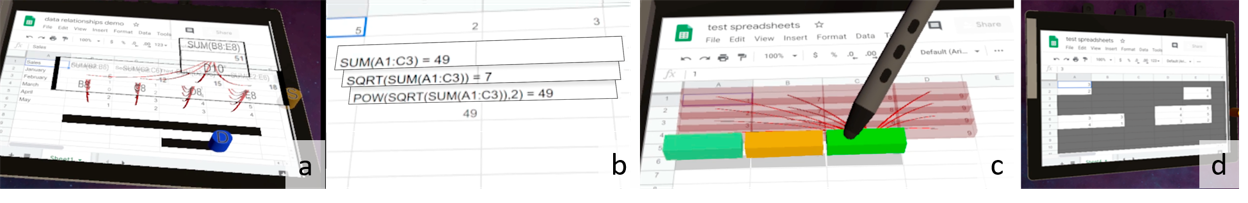}
  \caption{Cells are lifted up above their grid location and links visualize their dependency across two levels of computation (the black frame on the bottom show an option to cut the cells from the base layers) (a). While a cell of the grid displays the final result of the composed function, each layer may highlight the intermediate result of each added function (b). A cluster cell (bright green) is selected in a layer hovering above the main grid. The connected cells in the base layer are highlighted in red and also connected via red lines to the cluster cell (c) (shadows are used to better visualize the hovering layer in a monoscopic figure). A quick overview of used cells by masking the unused cells (d) (colored dark grey).}~\label{fig:cluster01}
\end{figure*}


Widely user adopted spreadsheet applications display functions' outcomes in a single cell, with no visualization of the cells that contributed to that calculation. While first-order dependencies (i.e.,~function parameters) may be visualized by coloring the cells when the user clicks on the result (function) cell, higher-level dependencies may not. Inspired by Shiozawa et al.~\cite{Shiozawa1999}'s visualization, the VR user can lift up a cell to see visual links to the dependent cells (Figure \ref{fig:cluster01}, a), as well as higher-order dependencies. Complementary, composed nested functions can also be shown using a stack visualization (Figure \ref{fig:cluster01}, b). Each layer represents one nesting function until the inner function is visible. 

\subsubsection{Cluster-cells}


Applying a function to a set of values, such as calculating a sum of cells, the user has to select each of the input cells while defining the function. This operation can be carried out by dragging the mouse over each range of cells or by typing their addresses. When dealing with a large amount of data, this selection process can quickly become exhausting, in particular when there is a need to re-select ranges for additional functions, or update ranges (for example, to remove outlines).


Our tool set introduces a concept of a {\em cluster-cell} that represents groups of cells as a unique entity. 
While each cluster-node has a unique cell location in the original grid for compatibility with 2D interactions, whenever the user enables a 3D display by tilting the pen from the screen surface toward the approximate height of a layer, or by selecting a button, the cluster-cells are visualized in a layer hovering above the main grid, corresponding geographically to the regular grid (Figure \ref{fig:cluster01}, c). Each cluster cell may correspond to one or more cells of similar semantics, originating in lower levels, generating a hierarchy of cluster-cells displayed in multiple overlay levels. Cluster-cells enable easy re-use of selections for different uses (such as different functions, or reuse as different axes in a graph) as well as a means to simplify the visualization of the structure of a spreadsheet by displaying large ranges using a single node. As cluster cells are elevated to their levels, the links to all the original cells they represent are displayed to the user.

There are several ways to define a cluster-cell. A data-to-cell approach requires the user to select input cells from lower levels, either regular cells or preceding cluster cells. They may lie in a continuous range, selected with a single swap of the pen, or be a set of disconnected cells, selected while continuously pressing a pen button. Lifting the pen tip to a higher layer, or selecting from an in-place menu, will generate a cluster-cell and allow the user to label it. Additionally, the user may use a cell-to-data approach and select an existing cluster-cell and add additional input cells by dragging a link (while pressing the pen's button) down to a lower level and selecting one or more cells. Releasing the button attaches the selected cells to the cluster-cell. Input cells of choice may be removed from the cluster by dragging them to the trash bin widget (seen in Figure \ref{fig:functioncreation}, h).

While displaying overlay layers, unused cells (cells in an overlay grid that have no assigned value) are displayed as transparent to increase visibility of lower layers. It is also possible to color unused cells of the original layer, to allow for a quick overview of the sheet content (Figure \ref{fig:cluster01}, d).



\subsection{Pen-based on and above Surface Interaction}



\begin{figure*}[tb!]
  \centering
  \includegraphics[width=2\columnwidth]{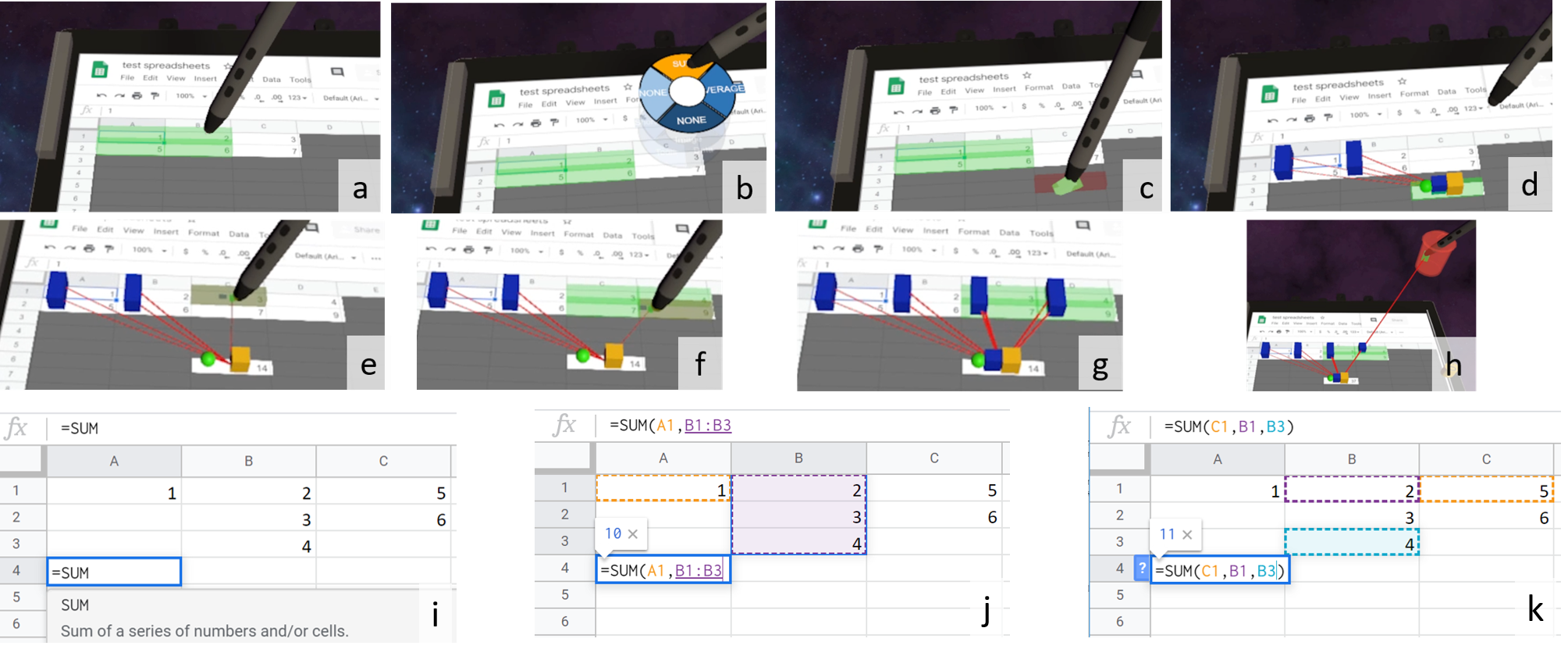}
  \caption{Top row: Adding a function workflow: selecting source cells (a), choosing a function from an in-place pie menu (b), selecting a cell for the function (c). Visualizing a function with links to it's source cells (d). The function may be re-positioned by moving the yellow button. Middle row: Adding and removing a function's source cells. Adding cells can be done by selecting with the pen (e), dragging to select a range of cells (f) and lifting the pen and drawing a red link to the function node (g) (inspired by the operation of a legacy telephone operator). Each source node displays a blue box, which can be dragged toward the garbage bin widget to dissociated them from the function (h). Bottom row: Example of creating a function workflow in common spreadsheet application. A target cell (A4) is selected (i), and a function is specified. Source cells are specified as function parameters individually (A1) or as a range (B1:B3) (j). Updating source cells using text entry (k). 
  } ~\label{fig:functioncreation}
\end{figure*}

Prior to presenting our combined in-air and touch techniques, let us consider function creation and editing using standard 2D spreadsheet software. 
Defining a function in applications such as Google Sheets (Figure \ref{fig:functioncreation}, bottom row) requires selecting a target cell for the function result (A4) and then specifying a function from either a menu or by typing the function name (Figure \ref{fig:functioncreation} i). The source cells are selected using a combination of pointing with the dominant-hand and simultaneously pressing a modifier-key on a physical keyboard (typically \texttt{CTRL}) with the non-dominant hand or by typing a list of cell ranges. When using finger touch or pen only, for example, when there is no keyboard or while holding the tablet, the duration for conducting this procedure becomes substantially longer (see our performance indication in Section \ref{sec:performance}). Specifically, selecting disjoint ranges of cells requires (potentially multiple) switches between selecting source cells, and entering delimiter signs (commas) on a virtual keyboard. 
Any editing of parameters of a function requires text editing, which is sometimes nontrivial (Figure \ref{fig:functioncreation} h). For example, removing a selected cell from a range of cells, (B2 from the range B1:B3), the range needs to be removed and individual cells need to be added again. Furthermore, re-using ranges for multiple uses, such as different functions, usage in graphs, and so on, requires re-selecting, or retyping the ranges.



Based on these observations, we designed a modified workflow combining touch and in-air interaction, which makes use of contextual hierarchical pie menus, visualizing and editing dependency links between cells (using a {\em telephone-operator metaphor}), and the use of cluster-cells.

\subsubsection{Menu Interaction}

There has been considerable effort in designing graphical menus  \cite{raymond1986survey, dachselt2007three, bailly2016visual}. Prior work has investigated the design and use of marking menus in VR using various input modalities such as a phantom \cite{komerska2004study}, fingers \cite{kulshreshth2014exploring, lim2019evaluation}, hands \cite{davis2016depth}, gaze \cite{pollock2019countmarks} or controllers with six degrees of freedom input \cite{gebhardt2013extended,  monteiro2019comparison} and techniques such as selection through ray-casting \cite{jeong2016ergonomic}, crossing \cite{tu2019crossing} or other gestures \cite{white2009interaction}. This prior body of work envisions the user pointing toward menus that are floating in front of the user in mid-air, spreading options apart, avoiding occlusions, such as overlapping menus, in order to enable easy mid-air selections. This paper, on the other hand, envisions the user working with hands supported by the tablet screen in a very similar fashion to 2D usage.

Inspired by Gebhardt et al.~\cite{gebhardt2013extended}, we designed an in-place hierarchical pie-menu that is operated using pen-based interaction while the hand is supported by the surface of the tablet and tilts the pen, lifting its tip to reach menus hovering above the tablet (Figure \ref{fig:menuschartshandles} a--c). While most of the motions are done by the fingers holding the pen, the hand remains static, reducing fatigue. A menu is invoked in-place \hl{using button press on the pen}, enabling the user to create a function or chart and place it in a target cell in a continuous motion (Figure \ref{fig:menuschartshandles} a--c). The user can confirm the final menu entry by button press. Also, the user may retract a menu by simply lowering the pen tip.

An additional arc-menu, on the lower right corner of the tablet, controls the display of meta-features. The arc shape of the menu allows for easy access to entries while the wrist is supported by the tablet, or by a thumb of a hand holding the tablet.





\subsubsection{Adding and updating functions}

While our prototype supports the 2D adding functions workflow, it also enables a new workflow that uses contextual menus: 1) select a target cell for the function; 2) raise the pen and select a function in the hovering hierarchical pie-menu; and 3) add source cells. Besides the above {\em function-to-source} workflow, the system also supports the inverse order {\em source-to-function}: First selecting source-cells, then selecting a function from a menu, and finally storing the function to a target cell in the grid (Figure \ref{fig:functioncreation}, a--d).
Source-to-function enables visualizing results of the function or, for example, a chart content while positioning it in its grid location.
Adding source cells to existing functions can be done in a similar way, by first selecting them and then drawing a link to the function node. (Figure \ref{fig:functioncreation}, e).

To avoid visual clutter, visualization can be toggled individually for each function, see Figure \ref{fig:menuschartshandles}, bottom row.

\subsubsection{Interacting with Charts}


The workflow for the creation of a chart (e.g., a bar chart) within a spreadsheet is very similar to creating a function cell. The user uses a pen to select source cells, then tilt the pen up to select a CHART option from the in-place menu, and completes the motion by selecting the location of the top left corner of the chart on the grid, dragging the pen to define the size of the chart. In this {\em source-to-chart} flow it is possible to view the chart being rendered while the pen defines this area. The entire workflow can be carried out with a single fluid motion of the pen. The created chart can be easily modified or resized afterward. 

\begin{figure*}[t!]
  \centering
  \includegraphics[width=2\columnwidth]{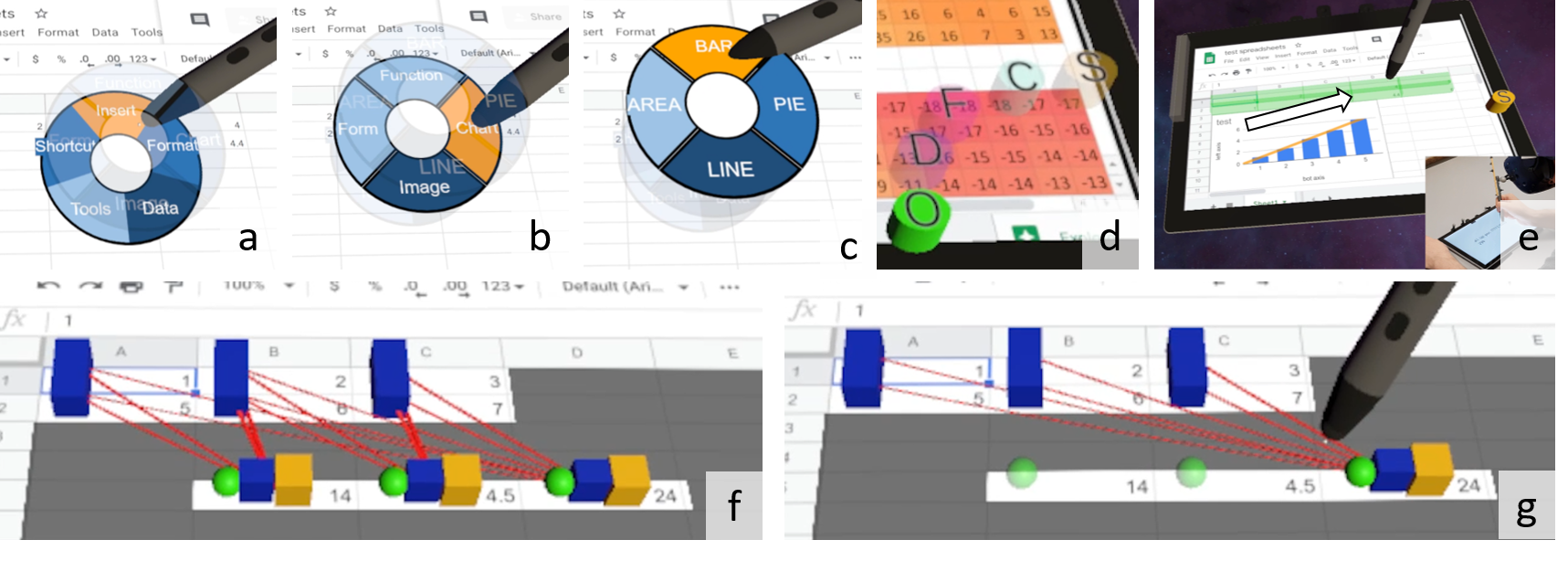}
  \caption{Top row: the hierarchical pie menu is invoked in-place and navigated from bottom to top through in-air movements of the pen (a-c). An arc menu on the corner of the tablet controls visualization options (activated - solid color, non activated -  semi-transparent) (d). Adding a trend-line to a chart through an in-air stroke gesture (e).  Bottom row: Three functions with their respective links to their source cells highlighted (f).  Green spherical buttons toggle displays of links per function. Here only the center function links display is enabled (g).  Abbreviations: O: overview window, D: dependency links, F: nested functions, C: cluster-nodes. S: additional sheets (tabs).
  }~\label{fig:menuschartshandles}
\end{figure*}

Furthermore, the user can modify the chart with in-air gestures such as adding a trend-line, by an in-air stroke of the pen tip over the chart (Figure \ref{fig:menuschartshandles}, e). Other regression models could be added, for example, a polynomial fit can be coupled with a curved in-air gesture.

\section{Implementation}
The described techniques were implemented in a VE using a Unity game engine\footnote{https://unity.com/ Last access May 18, 2020}, commodity PC hardware, and an HTC Vive Pro HMD. The system includes a fully interactive spreadsheet implementation based on Google Sheets\footnote{https://www.google.com/sheets/about/ Last access May 18, 2020}. A separate Microsoft Surface Pro 4 tablet was used to sense pen input data including touches and button press events and sent them to the Unity application via UDP network communication. Both the tablet and the pen were spatially tracked via an OptiTrack motion-tracking system, and were rendered in the Unity virtual environment by similar sized 3D models.

Google Sheets website pages are rendered inside Unity using embedded instances of a Chromium browser engine, provided by the ZFBrowser Embedded Browser Unity plugin\footnote{https://assetstore.unity.com/packages/tools/gui/embedded-browser-55459 Last access May 18, 2020}. ZFBrowser renders the contents of opened web pages into a texture with the same resolution as the physical \hl{Surface Pro 4} to achieve equivalent scaling of rendered HTML elements between the physical and virtual notebook. This texture is then mapped onto the screen of the virtual notebook, and touch interaction coordinates received via UDP from the notebook can be mapped onto this texture in a one-to-one fashion. Optionally, input coordinates can be normalized to a [0,1] range in x and y directions by dividing by the physical screen width and height to remap the input onto output surfaces of arbitrary dimensions.

While interacting with the sheets web interface, there is no way to define or change functions and charts in one step, so we used the Google Sheets cloud API to apply cell transformations. Since the Google Sheets API exposes no functionality to track client-side interactions, operations of users with the web page were tracked inside Unity. 
In particular, tracking of cell selection was implemented by constructing virtual-cells in the Unity space using oriented bounding boxes, and spatially position them in their corresponding places to fit the spreadsheet texture. Tracking of the pen and the Unity collision detection mechanism is used to detect whether the pen tip lies inside a certain cell\footnote{https://docs.unity3d.com/ScriptReference/Collider.OnTriggerEnter.html Last access May 18, 2020}. 

Enabling a display of only used cells in overlay layers was implemented using a custom alpha masking technique. First, to identify empty cells we use the Google Sheets API to retrieve the cell values for each cell in a displayed spreadsheet. Then, a virtual camera renders cells that should appear transparent into a separate monochrome texture, called an alpha mask. The cells are rendered into this texture such that the  alpha mask is registered with the rendered browser texture. A custom shader uses the corresponding alpha mask values and rendered browser pixels transparent accordingly. 

To facilitate further research and development of spreadsheet interaction within VR, our code is available under https://gitlab.com/mixedrealitylab/spreadsheetvr.

\section{Evaluation}
We validated our prototype through an online survey and by gathering performance data from expert users. Both evaluations are described next. 

\subsection{Video-based Survey}

While we used informal user tests throughout the iterative design process of the described techniques, we gathered further user feedback on the potential usefulness and attractiveness of the techniques through a video-based online survey. 

\subsubsection{Design and Procedure}
We ran an online experiment \hl{using a within-subjects design}, in which the individual techniques were presented to users as video prototypes. Please note that while users were not able to try the techniques out for themselves as interactive prototypes, collecting user feedback based on video prototypes is an established practice (e.g., \cite{chen2014duet, dhillon2011visual, syrdal2008video}).
The videos showed the following twelve techniques: 
creating functions (marked as CF), 
manipulating functions (adding, removing data cells) (MF), 
creating cluster-nodes (CC), 
selection across multiple tabs using gaze-and-touch interaction (SE), 
selection across multiple tabs using mid-air interaction (SA), 
Display of an overview window of the spreadsheet (OV), 
extended view of a spreadsheet beyond the tablet's view-port (EV), 
visualization of cell dependencies (DV), 
visualization of nested functions (NF), 
masking unused cells, (MC) 
chart interaction (CT) 
as well as a close-up of the hierarchical pie-menu interaction (PM) (which was also used as integral part of other techniques).

For each presented technique, users were asked to rate its usefulness, how easy-to-use it looks, if they would recommend it to a friend and how they would rate the technique overall. Further, participants were free to add further comments on each technique.

Please note, that the presentation of the techniques was not counterbalanced. While this could lead to ordering effects, some techniques were based on using prior presented techniques (e.g. adding or manipulating functions). With counterbalancing in place, this could have limited participants' comprehension of the respective techniques. 

After all techniques were presented, the participants were asked to select one technique that they liked best overall, as well as one technique, that they liked least overall. They were then asked to state reasons for their respective choices. 

Overall, the evaluation took about 30 minutes per participant. No compensation was paid to participants.

\subsubsection{Participants}
We recruited 18 participants (3 female, 15 male, mean age 31 years, sd = 8.39). 
Four participants reported a very high level of experience with virtual reality, eight stated a high experience level, four said that they were moderately experienced and two indicated little experience. Three participants indicated a very high level of experience with pen-based systems, five stated a high experience level, four reported moderate experience, three indicated little experience and three had no experience at all. Six participants reported a very high level of experience with spreadsheet applications, six reported a high level of experience, five stated they were moderately experienced and one reported little experience. Three participants reported that they very frequently use spreadsheet applications on mobile devices. Two stated that they used them frequently, three participants use them sometimes, seven rarely and three never.


\subsubsection{Results}

\begin{figure*}[htb!]
  \centering
  \includegraphics[width=2\columnwidth]{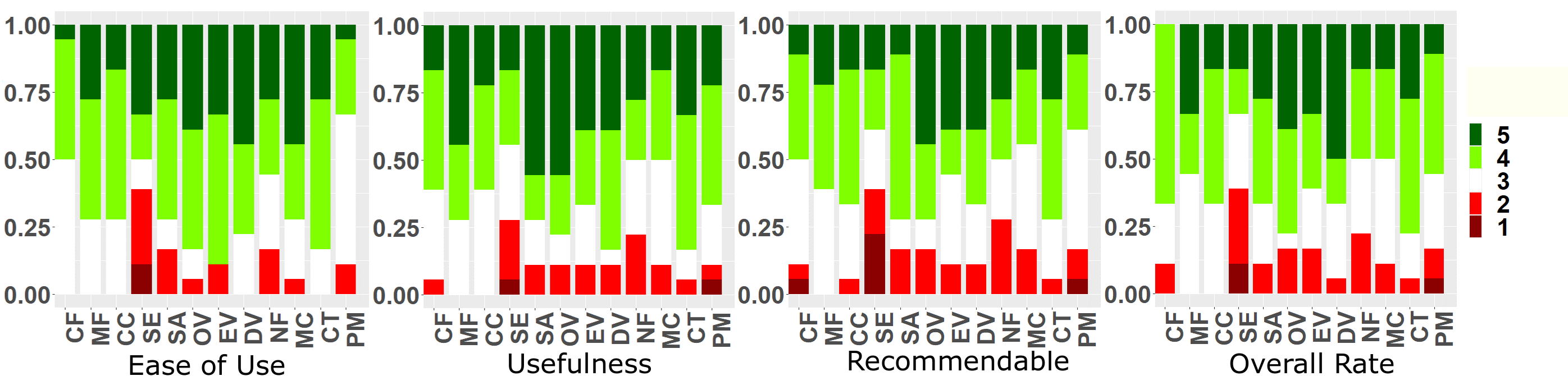}
  \caption{Rating overview on ease of use, usefulness, recommendability, and overall rating on a five-item Likert scale (1: strongly disagree, 5 strongly agree). Techniques: CF: creating functions, MF: manipulating functions, CC: clustering cells, SE: gaze+touch interaction across multiple tabs, SA: in-air interaction across multiple tabs, OV: overview visulization, EV: extended viewport, DV: dependency visualization, NF: visualization of nested functions, MC: masking of unused cells, CT: Adding a trendline to a chart, PM: hierarchical pie menu. The \hl{y-axes depicts} the number of participants in percent (1.0  = 100\% of participants).}~\label{fig:ratings}
\end{figure*}

The results for the user ratings are depicted in Figure \ref{fig:ratings}. Friedman tests revealed significant differences, between conditions, see Table \ref{tab:resultsratings}. However, post-hoc tests using Wilcoxon signed rank tests with Bonferroni correction, did not reveal significant pairwise differences. 

\begin{figure}[htb!]
  \centering
  \includegraphics[width=0.9\columnwidth]{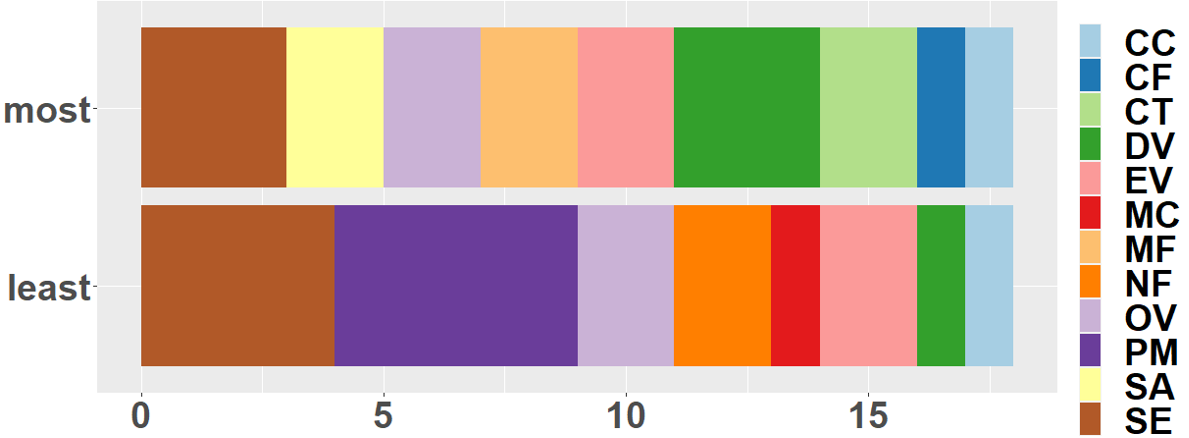}
  \caption{Most and least preferred techniques as stated by the participants. Techniques: CF: creating functions, MF: manipulating functions, CC: clustering cells, SE: gaze+touch interaction across multiple sheets, SA: in-air interaction across multiple sheets, OV: overview visulization, EV: extended viewport, DV: dependency visualization, NF: visualization of nested functions, MC: masking of unused cells, CT: Adding a trendline to a chart, PM: hierarchical pie. The x-axis depicts the number of participants in percent (1.0  = 100\% of participants) menu.}~\label{fig:preferences}
  \vspace{-0.5cm}
\end{figure}

We also asked users to identify their most preferred and least preferred function, the results are depicted in Figure \ref{fig:preferences}. 



Besides, rating the techniques, we asked users to comment on the individual techniques.

\begin{table}[b!]
\centering
\caption{Results of  Friedman tests with regards to the four different categories (usefulness, ease of use, recommendation and overall rating)}
\begin{tabular}{|c|c|c|c|}
\hline 
 & $\chi^2(11)$ &$ p$ & $W$ \\ 
\hline 
Usefulness &$ 26.0$ & $.006$ & $0.13$ \\ 
\hline 
Ease of use & $31.6$ & $.0008$ & $0.16$ \\ 
\hline 
Recommendation & $25.6$ & $.007$ & $0.13$ \\ 
\hline 
Overall rating & $26.2$ & $.006$ &$ 0.13$ \\ 
\hline 
\end{tabular} 

\label{tab:resultsratings}
\end{table}



{\bf creating functions}: Participant 7 (P7) stated that  "Looks like it would be fun to work with". as well as "working with symbols [boxes and links] is much quicker than working with large numbers or functions." P14 stated that "allows to see where the values are coming from."
While users stated that selecting functions from pie-menus are "pretty straightforward" to use, this method "may not work for the many functions not on the menu" (P8).
P16 was concerned about "the high degree of motor control that the technique seems to require" and P17 noted that 3D cells "are potentially distracting and may occlude important content (in contrast to the more typical use of border colours to indicate a cell range)"

{\bf manipulating functions}: P8 mentioned that adding source cells and toggling the links "appears to be easy". P14 stated that buttons on the pen "would be better for activating the different modalities." P4 mentioned that this technique "seems useful and faster than the "normal" approach in excel." P13 stated that the technique is "easy to understand. shows links. could be used in normal excel." Three users stated that deleting is "too complicated" due to added dragging toward the trash can." P14 suggested a possible solution "to show the bin closer to the hand once a cell is activated".
Three participants proposed to using additional pen buttons to overload functions (e.g., switching between adding and deleting of cells). 

{\bf cluster cells}: P8 mentioned "the technique is useful. I had not seen aliases for cell ranges before", that the motion "seems easy", but also wondered if it would be "frequently triggered accidentally".
P14 said that the technique is "really good" and that "creating on the fly shortcuts or templates would reduce a lot of time" as it's "very common in excel to do the same operations for lots of combinations of columns, where the same function is copied and modified over and over." P9 mentioned "It makes it easy to group data without having to use multiple sheets and thus reduces complexity."

{\bf interaction across multiple sheets using gaze-and-touch}: P7 stated the technique seams "very easy" as well as "I can get a better overview of my different worksheets and select them very quickly. This technique would be a great relief in my daily work." P10 stated "Great way to visualize the connection to other sheets and I assume it works better with a larger number of sheets than the in-air technique and therefore I prefer it." P18 stated "It made something that is difficult to do today much easier."

On the contrary, three users commented that head motions seem unnecessary large. P8 suggested that a better layout of neighboring tabs may allow for smaller head motions. 

{\bf interaction across multiple sheets using in-air interactions}: P8 suggested to extend the technique to far away tabs, by "magnify[ing]" them once the user reaches out for them.
P17 liked the ability to work with two sheets side-by-side. "Main issue here is that you are forced to operate at greater distance on second sheet which will negatively impact legibility and selection accuracy."

{\bf overview visualization}: P14 thought it was "really good and make[s] more sense for navigating spreadsheets rather than scrolling". P17 called it "a great idea" in particular for data sets. P5 called the technique "very handy to navigate large data sheets" and P16 liked "the separation of views".

{\bf extended view beyond the tablet view-port}: P16 liked "the spreadsheet breaks out of the boundaries" of the tablet and P17 called "leveraging [the] expanded display region and orientation freedom" a great idea. P16 also mentioned a possible way for improvement: "perhaps using the same approach to maps (where one zooms out when moving across regions) might make this technique better."

{\bf dependency visualization}: P7 mentioned that the technique seems "very useful to check quickly the correctness" of spreadsheets calculations. P8 liked the idea of using the depth to separate out layers of information. P3 called it "extremely useful" for large sheets. P8 stated that "Tracking down cell dependencies in Excel is one of the banes of my existence. The technique is easy to use and presents information in an intuitive way. Also, I feel that it is suitable to VR and leverages the affordances of VR." P14 mentioned "This can be very useful to check and verify that certain functions are correctly set up."

{\bf masking unused cells}, P9 thought masking unused cells "makes it easier to find important content".

{\bf chart interaction}: P12 said this technique seems "easy and natural to use."

{\bf hierarchical in-place pie menus}: P7 was reminded of "drawing with the essential fine motor skills." P17 thought it was "cool" but warns of a potential problem as "is that place in hierarchy and path to current menu is partially hidden." P12 mentioned the menu seems "quite confusing and hard to manipulate" and P16 was "not a fan of mid-air stacks as a selection mechanism".

\subsection{Indicative Human Performance Potential}
\label{sec:performance}
We had five expert users (four male, one female, mean age 28.4 years sd = 5.13), who trained to be proficient in executing common tasks in spreadsheet software. 


\hl{The lab-based experiment had two independent variables. The first one was the interaction technique which was either \textsc{VR}, \textsc{tablet and pen} or \textsc{keyboard and touchpad}. The second one was the task which included creating a function (CF), adding individual cells to a function (AIC), adding a range to a function (AR), remove single cells from a function (RC), reusing cells when creating a function (RE), adding a chart (AC) and adding a trendline (AT) and adding individual cells from another tab (AICS).
The participants executed all of these tasks and repeated them five times with each interaction technique} (see video in the supplementary material for footage of a selected expert user). \hl{We used counterbalancing, for the tasks as well as for the interaction techniques, to mitigate learning effects}. Please note that we used simple tasks on purpose, as more complex tasks are typically composed of those atomic parts, even though this might be unfavorable for the VR techniques. For example, for reusing cells when creating a function, participants could reuse cells by copying a cell range initially and pasting it into several functions using \textsc{tablet and pen} as well as \textsc{keyboard and touchpad} techniques. This process of copying once and pasting multiple times might not work \hl{efficiently in scenarios, in which several cell ranges and multiple functions are used. Here, copy and paste could result in higher coordination efforts (i.e. to avoid copying the same range multiple times, a range first has to be copied in all functions, then the next range should be copied and pasted).}
Besides \textsc{tablet and pen} we included \textsc{keyboard and touchpad} interaction as an additional reference point, when users would have access to physical keyboards, such as on notebooks or on selected slates with detachable keyboard. \hl{When using text entry in either \textsc{tablet and pen} or \textsc{keyboard and touchpad}, closing brackets could be ommitted to speed up text entry. Please also note, that in the \textsc{VR} condition, users did not need to explicitly enter text. We tried to design the best possible techniques in VR and indeed found that they do not require text entry for the tasks at hand.}

The task completion times were computed as the duration between the first selection of a relevant cell (e.g., for entering a function or selecting source cells) until the result was displayed (e.g., the function result, a chart). This way, initial travel times from various potential starting points have not been included. Overall, the data collection took around 90 minutes per participant. No compensation was paid.

The results for task completion time are depicted in Table \ref{tab:TCT}. Please note, that these times should solely indicate the performance that can be achieved with sufficient training. While we depict results of repeated measures analysis of variance (RM-ANOVA, data was tested for normality with the Shapiro-Wilk normality test) along with Bonferroni adjusted post-hoc tests (initial significance level $\alpha = .05$), the results are solely indicative, both due to the small sample size and the lack of integrating users with a more representative background. 
\begin{table}[t]

\centering
\caption{Task completion time in seconds (average, standard deviation in parenthesis below) for the different applications and the three interfaces virtual reality (VR), tablet and pen (TP) as well as keyboard and touchpad (KT). Tasks (T): creating a function (CF), adding individual cells to a function (AIC), adding a range to a function (AR), remove single cells from a function (RC), reusing cells when creating a function (RE), adding a chart (AC) and adding individual cells from another sheet (AIS). For AICS, both gaze+touch (G+T) as well as in-air selection (IA). Pairwise significant differences at an initial significance level of $\alpha = .05$  are indicated in the * column.}
\small
\begin{tabular}{|c|c|c|c|c|c|c|c|c|c|}
\hline 
\textbf{T} & \multicolumn{2}{|c|}{\textbf{VR}} & \textbf{TP} & \textbf{KT} & \textbf{$F_{(2,8)}$} & \textbf{p} & ${\eta^2_p}$ & *\\ 
\hline 
CF & \multicolumn{2}{|c|}{$6.62$} & $12.7 $ & $4.96$ & $15.2$ & $.002$ & $.79$ & VR-TP\\ 
 & \multicolumn{2}{|c|}{$(1.30)$} & $(3.62)$ & $(1.49)$ &  &  &  & KT-TP\\  
 \hline
AIC & \multicolumn{2}{|c|}{$2.70$} & $6.70 $ & $3.07$ & $29.5$ & $<.001$ & $.88$ & VR-TP\\ 
 & \multicolumn{2}{|c|}{$(0.59)$} & $(1.11)$ & $(0.47)$ &  &  &  & KT-TP\\ 
\hline 
AR & \multicolumn{2}{|c|}{$3.53$} & $10.4 $ & $3.93$ & $40.2$ & $<.001$ & $.91$ & VR-TP\\ 
 & \multicolumn{2}{|c|}{$(0.79$} & $(1.95)$ & $(0.81)$ &  &  &  & KT-TP\\ 
 \hline 
RC & \multicolumn{2}{|c|}{$2.21$} & $12.3 $ & $6.75$ & $18.3$ & $.001$ & $.82$ & VR-TP\\ 
 & \multicolumn{2}{|c|}{$(0.61)$} & $(4.86)$ & $(1.65)$ &  &  &  & KT-TP\\  
 \hline
RE & \multicolumn{2}{|c|}{$5.71$} & $10.6 $ & $3.06$ & $11.0$ & $.005$ & $.73$ & VR-TP\\ 
 & \multicolumn{2}{|c|}{$(1.97)$} & $(3.26)$ & $(1.53)$ &  &  &  & KT-TP\\  
 \hline
AC & \multicolumn{2}{|c|}{$9.03$} & $7.17 $ & $4.93$ & $3.97$ & $.06$ & $.50$ &\\ 
 & \multicolumn{2}{|c|}{$(1.74)$} & $(2.92)$ & $(0.74)$ &  &  &  &\\  
 \hline 
 & IA & G+T & \multicolumn{2}{|c|}{} & \textbf{$F_{(3,12)}$} & \multicolumn{3}{|c|}{}\\ 
\hline 
AIS & $3.94$ & $6.77$ & $8.39$ & $5.72$ & $4.22$ & $.03$ & $0.51$ & IA-TP\\ 
  & $(0.99)$ &  $(2.30)$ & $(2.56)$ & $(0.60)$ &  &  &  & \\ 
\hline 
\end{tabular} 

\label{tab:TCT}
\end{table}

\section{Discussion}

We presented a first step in investigating the use of VR to improve the user experience of spreadsheet work on the go. We use VR to extend the user display space and enable better visualization and interaction of spreadsheets while simultaneously maintaining a small interaction-space around the tablet. We also want to maximize users' familiarity with common 2D spreadsheet workflows and use the support of the tablet for the user's hands to enable long period of work with less fatigue.

This work has raised a set of principles, such as a separation between the input-space and the data/display-space, compatibility of representations between the VR and 2D workflows and limiting the input around familiar touchscreen devices. While we expect more interaction techniques to be suggested in the future, we also expect these principles to be maintained.

Our indicative evaluations revealed that the techniques, were mostly deemed usable and useful by participants in a video-based survey. No technique was clearly preferred by the majority of users. However, the feedback users provided on individual functions, highlighted both opportunities for further improvement of the techniques as well as potential challenges when running a video-based evaluation. For example, gaze-and-touch may look confusing when observed in a video, as indicated by participant ratings, due to the fast apparent head motions. In contrast, when developing this technique, in our informal tests, we found the technique to be very comfortable when wearing an HMD and using it in-situ. Similarly, several users were concerned with in-air interaction with hierarchical pie-menus as well as with visual clutter. Again, when using 'dogfood' prototype in internal testing, we found the technique to be comfortable and efficient due to the support of a resting wrist on the table or tablet, with little rotations of the pen for menu selection (instead of lifting up the entire hand or arm). The transparency of the non selected layers, which may look cluttered, enables a glance at the next level of menus without covering the underlying opaque layer when seen through a stereo HMD. 

The indicative human performance evaluation with expert users further revealed that the proposed techniques can be efficient to use. For several base functions such as creating functions, adding (individual or multiple) cells, removing or reusing cells, VR was significantly faster than tablet and pen interaction. However, please note that these evaluations should be seen as indicative. In future work, walk-up-and-use usability and performance should be tested with non-expert users in interactive sessions. Further, the performance of compound tasks (complementary to the atomic tasks used in the expert evaluation) could be further studied. Nonetheless, the expert evaluation provided evidence that the system can be usable and be used efficiently and the video-based evaluation revealed some preliminary evidence of wider user interest in this new approach to mobile spreadsheet interaction.

\hl{The choices of the experimental designs were influenced and limited by the COVID-19 restrictions put in place at our university. Therefore, the results of the studies have to be interpreted with caution and further studies should be conducted.}

The current prototype was implemented in a lab, using an external OptiTrack device for tracking the pen and the tablet to achieve the best spatial tracking we can get.  However, similar tracking capabilities are becoming accessible for mobile settings\hl{, even though the accuracy of current generation mobile tracking systems is substantially lower compared to dedicated outside-in tracking systems like OptiTrack} \cite{schneider2020accuracy}. The use of head mounted cameras on HMDs has become prevalent for both inside-out tracking, and for streaming outside video for video-based augmented reality applications. Using the video stream from those cameras, the tablet and the pen can be tracked (the tablet screen may display fiducials for pose estimation as it is not seen by the user). Finger tracking technologies are appearing in recent HMDs, such as Hololens 2.0, Oculus Quest, Vive Pro and others. This would also enable to study spreadsheet interactions in various mobile real-world scenarios.

Further, the examples in this paper were executed using a pen, as this input device combines several advantages: it enables fine accuracy of touch input, which can be used for text entry with hand writing. This is easy to do in-place and using one hand, while the non-dominant-hand holds the tablet. It is also easy to track the pen by adding a tracker at it's back tip, and it has buttons that can be used to control operations, such as selecting multiple cells one after the other. However, our techniques can be extended to be used with other input modalities, such as a finger touch, where the non-dominant-hand touch may simulate a button, a soft keyboard is used for typing, and the raising index finger replaces pen tilting. If space allows, a mouse and a keyboard may also be used, where the mouse pointer is the location of the pen, and the mouse wheel or a touch sensitive area on the mouse may be used to tilt the pen upward.  

In most of the figures of this paper, we rendered the display space parallel to the surface of the tablet. Such a display is easy to understand, as it is clear where the tablet and the user's pen are. However, in many applications, where the users have to work in very limited places, such as an airplane seat, we envision users moving the display space to a vertical plane ahead of them (such as in Figure \ref{fig:extended-view}), while the hands are still supported on the tablet. Every movement of the pen or the hands is represented as a corresponding movement of the virtual pen or hands in the display space (as explored by Grubert et. al.~\cite{grubert2018text}). Finally, the proposed interaction techniques could be explored in augmented reality (AR) using optical see-through (OST) HMDs. While the field of view of immersive VR HMDs is typically substantially larger compared to OST HMDs, most VR HMDs do not match the output resolution of a tablet's screen. Hence, in our work, the number of cells being concurrently legible was smaller than on a typical 4K tablet screen, \hl{potentially leading to an increased need for navigating between cells that are far apart}. OST HMDs could potentially be leveraged to show additional information about spreadsheets while still allowing use of a high resolution screen.




\section{Conclusions and Future Work}

Within this work, we presented a toolset for enhancing mobile spreadsheet interaction on tablets using immersive VR HMDs and pen-based input. We proposed to use the space around the tablet for enhanced visualization of spreadsheet data and presented a set of interaction techniques for around and above tablet visualization of spreadsheet-related information, for example, to allow previewing cells beyond the bounds of the physical touchscreen from the same or adjacent tabs or to uncover hidden dependencies between cells. Combining the input capabilities of precise on-surface, pen-based interaction along with spatial sensing around the tablet, we demonstrated techniques for efficient creation and editing of spreadsheets functions. Our indicative studies showed the feasibility of our approach. In future work, our initial explorations and evaluations should be extended to cover walk-up-and-use scenarios, more complex compound tasks. Further, we are interested in exploring how VR spreadsheet interactions could work in collaborative scenarios. For instance, when interacting with multiple sheets, being able to visualize the hands of other users working on different sheets for awareness or to facilitate collaboratively working with them. We are also interesting in exploring how users could edit or modify a spreadsheet's content in VR using the pen, such as using handwriting in the cells, or using or adjusting values in cells with real-time updates of the charts by using relative movements, for example, similar to Pfeuffer et. al~\cite{pfeuffer2017thumb} but in VR.  This work has been focusing on pen and touch interactions, but there might be situations where users do not have a pen or prefer using touch only, such as in an airplane where they could be worried to lose the pen. To that end, we are interested in studying a variant of the spreadsheet VR experience with only touch and compare the usability and performance with pen and touch. We are also contemplating the possibility to extend our research to explore heads-up experiences by using pen and touch to manipulate spreadsheets indirectly while having the visuals situated in front of the user. \hl{Further, in future tasks that also require text entry in VR could be studied. Also, while for our performance study we reduced the number of concurrently visible cells in favor of a comparable field of view between the virtual and physical screen, one could also change the field of view of the virtual screen to match a higher number of visible cells typically concurrently visible on physical 4K screens.} Finally, the use of augmented keyboards~\cite{schneider2019reconviguration} for supporting spreadsheet interaction should be further investigated, for example, by re-purposing keys for navigating between data types (e.g., to jump to next numeric or character).


\balance{}

\bibliographystyle{abbrv-doi}

\bibliography{spreadsheet.bib}

\begin{thebibliography}{10}

\bibitem{bailly2016visual}
G.~Bailly, E.~Lecolinet, and L.~Nigay.
\newblock Visual menu techniques.
\newblock {\em ACM Computing Surveys (CSUR)}, 49(4):1--41, 2016.

\bibitem{batmaz2020precision}
A.~U. Batmaz, A.~K. Mutasim, and W.~Stuerzlinger.
\newblock Precision vs. power grip: A comparison of pen grip styles for
  selection in virtual reality.

\bibitem{Biener2020Breaking}
V.~Biener, D.~Schneider, T.~Gesslein, A.~Otte, S.~Kuth, P.~O. Kristensson,
  E.~Ofek, M.~Pahud, and J.~Grubert.
\newblock Breaking the screen: Interaction across touchscreen boundaries in
  virtual reality for mobile knowledge workers.
\newblock In {\em IEEE transactions on visualization and computer graphics},
  2020.

\bibitem{bier1993}
E.~A. Bier, M.~Stone, K.~Pier, W.~Buxton, and D.~T.
\newblock Toolglass and magic lenses: the see-through interface.
\newblock pp. 73--80. ACM, 1993.

\bibitem{birch2018future}
D.~Birch, D.~Lyford-Smith, and Y.~Guo.
\newblock The future of spreadsheets in the big data era.
\newblock {\em arXiv preprint arXiv:1801.10231}, 2018.

\bibitem{blackwell2003notational}
A.~Blackwell and T.~Green.
\newblock Notational systems--the cognitive dimensions of notations framework.
\newblock {\em HCI models, theories, and frameworks: toward an
  interdisciplinary science. Morgan Kaufmann}, 2003.

\bibitem{brandl2008combining}
P.~Brandl, C.~Forlines, D.~Wigdor, M.~Haller, and C.~Shen.
\newblock Combining and measuring the benefits of bimanual pen and direct-touch
  interaction on horizontal interfaces.
\newblock In {\em Proceedings of the Working Conference on Advanced Visual
  Interfaces}, AVI ’08, p. 154–161. Association for Computing Machinery,
  New York, NY, USA, 2008. doi: {{%
10\hspace{.1pt}\discretionary{.}{%
}{.}\hspace{.4pt}1145\discretionary{/}{%
}{/}1385569\hspace{.1pt}\discretionary{.}{%
}{.}\hspace{.4pt}1385595}}


\bibitem{burnett2001forms}
M.~Burnett, J.~Atwood, R.~W. Djang, J.~Reichwein, H.~Gottfried, and S.~Yang.
\newblock Forms/3: A first-order visual language to explore the boundaries of
  the spreadsheet paradigm.
\newblock {\em Journal of functional programming}, 11(2):155--206, 2001.

\bibitem{buschel2018interaction}
W.~B{\"u}schel, J.~Chen, R.~Dachselt, S.~Drucker, T.~Dwyer, C.~G{\"o}rg,
  T.~Isenberg, A.~Kerren, C.~North, and W.~Stuerzlinger.
\newblock Interaction for immersive analytics.
\newblock In {\em Immersive Analytics}, pp. 95--138. Springer, 2018.

\bibitem{cami2018unimanual}
D.~Cami, F.~Matulic, R.~G. Calland, B.~Vogel, and D.~Vogel.
\newblock Unimanual pen+ touch input using variations of precision grip
  postures.
\newblock In {\em Proceedings of the 31st Annual ACM Symposium on User
  Interface Software and Technology}, pp. 825--837, 2018.

\bibitem{chambers2010struggling}
C.~Chambers and C.~Scaffidi.
\newblock Struggling to excel: A field study of challenges faced by spreadsheet
  users.
\newblock In {\em 2010 IEEE Symposium on Visual Languages and Human-Centric
  Computing}, pp. 187--194. IEEE, 2010.

\bibitem{chen2014duet}
X.~Chen, T.~Grossman, D.~J. Wigdor, and G.~Fitzmaurice.
\newblock Duet: exploring joint interactions on a smart phone and a smart
  watch.
\newblock In {\em Proceedings of the SIGCHI Conference on Human Factors in
  Computing Systems}, pp. 159--168, 2014.

\bibitem{chen2014air+}
X.~Chen, J.~Schwarz, C.~Harrison, J.~Mankoff, and S.~E. Hudson.
\newblock Air+ touch: interweaving touch \& in-air gestures.
\newblock In {\em Proceedings of the 27th annual ACM symposium on User
  interface software and technology}, pp. 519--525, 2014.

\bibitem{chintapalli2016comparative}
V.~V. Chintapalli, W.~Tao, Z.~Meng, K.~Zhang, J.~Kong, and Y.~Ge.
\newblock A comparative study of spreadsheet applications on mobile devices.
\newblock {\em Mobile Information Systems}, 2016, 2016.

\bibitem{dachselt2007three}
R.~Dachselt and A.~H{\"u}bner.
\newblock Three-dimensional menus: A survey and taxonomy.
\newblock {\em Computers \& Graphics}, 31(1):53--65, 2007.

\bibitem{davis2016depth}
M.~M. Davis, J.~L. Gabbard, D.~A. Bowman, and D.~Gracanin.
\newblock Depth-based 3d gesture multi-level radial menu for virtual object
  manipulation.
\newblock In {\em 2016 IEEE Virtual Reality (VR)}, pp. 169--170. IEEE, 2016.

\bibitem{dhillon2011visual}
B.~Dhillon, P.~Banach, R.~Kocielnik, J.~P. Emparanza, I.~Politis,
  A.~R^^1D8Fczewska, and P.~Markopoulos.
\newblock Visual fidelity of video prototypes and user feedback: a case study.
\newblock In {\em Proceedings of HCI 2011 The 25th BCS Conference on Human
  Computer Interaction 25}, pp. 139--144, 2011.

\bibitem{drachen2018games}
A.~Drachen, P.~Mirza-Babaei, and L.~E. Nacke.
\newblock {\em Games user research}.
\newblock Oxford University Press, 2018.

\bibitem{fellion2017flexstylus}
N.~Fellion, T.~Pietrzak, and A.~Girouard.
\newblock Flexstylus: Leveraging bend input for pen interaction.
\newblock In {\em Proceedings of the 30th Annual ACM Symposium on User
  Interface Software and Technology}, UIST ’17, p. 375–385. Association for
  Computing Machinery, New York, NY, USA, 2017. doi: {{%
10\hspace{.1pt}\discretionary{.}{%
}{.}\hspace{.4pt}1145\discretionary{/}{%
}{/}3126594\hspace{.1pt}\discretionary{.}{%
}{.}\hspace{.4pt}3126597}}


\bibitem{Fitzmaurice03trackingmenus}
G.~Fitzmaurice, A.~Khan, R.~Pieké, B.~Buxton, and G.~Kurtenbach.
\newblock Tracking menus.
\newblock In {\em UIST}, pp. 71--79. ACM Press, 2003.

\bibitem{flood2008evaluation}
D.~Flood, K.~M. Daid, F.~M. Caffery, and B.~Bishop.
\newblock Evaluation of an intelligent assistive technology for voice
  navigation of spreadsheets.
\newblock {\em arXiv preprint arXiv:0809.3571}, 2008.

\bibitem{flood2011systematic}
D.~Flood, R.~Harrison, C.~Martin, and K.~McDaid.
\newblock A systematic evaluation of mobile spreadsheet apps.
\newblock In {\em IADIS International Conference Interfaces and Human Computer
  Interaction}, pp. 1--8, 2011.

\bibitem{flood2011spreadsheets}
D.~Flood, R.~Harrison, and K.~McDaid.
\newblock Spreadsheets on the move: An evaluation of mobile spreadsheets.
\newblock {\em arXiv preprint arXiv:1112.4191}, 2011.

\bibitem{flood2011useful}
D.~Flood, R.~Harrison, and A.~Nosseir.
\newblock Useful but tedious: An evaluation of mobile spreadsheets.
\newblock In {\em PPIG}, p.~6. Citeseer, 2011.

\bibitem{gebhardt2013extended}
S.~Gebhardt, S.~Pick, F.~Leithold, B.~Hentschel, and T.~Kuhlen.
\newblock Extended pie menus for immersive virtual environments.
\newblock {\em IEEE transactions on visualization and computer graphics},
  19(4):644--651, 2013.

\bibitem{grasset2007mixed}
R.~Grasset, A.~Duenser, H.~Seichter, and M.~Billinghurst.
\newblock The mixed reality book: a new multimedia reading experience.
\newblock In {\em CHI'07 extended abstracts on Human factors in computing
  systems}, pp. 1953--1958. ACM, 2007.

\bibitem{grossmann2006hover}
T.~Grossman, K.~Hinckley, P.~Baudisch, M.~Agrawala, and R.~Balakrishnan.
\newblock Hover widgets: Using the tracking state to extend the capabilities of
  pen-operated devices.
\newblock In {\em Proceedings of the SIGCHI Conference on Human Factors in
  Computing Systems}, CHI ’06, p. 861–870. Association for Computing
  Machinery, New York, NY, USA, 2006. doi: {{%
10\hspace{.1pt}\discretionary{.}{%
}{.}\hspace{.4pt}1145\discretionary{/}{%
}{/}1124772\hspace{.1pt}\discretionary{.}{%
}{.}\hspace{.4pt}1124898}}


\bibitem{grubert2018office}
J.~Grubert, E.~Ofek, M.~Pahud, P.~O. Kristensson, F.~Steinicke, and C.~Sandor.
\newblock The office of the future: Virtual, portable, and global.
\newblock {\em IEEE computer graphics and applications}, 38(6):125--133, 2018.

\bibitem{grubert2018text}
J.~Grubert, L.~Witzani, E.~Ofek, M.~Pahud, M.~Kranz, and P.~O. Kristensson.
\newblock Text entry in immersive head-mounted display-based virtual reality
  using standard keyboards.
\newblock In {\em 2018 IEEE Conference on Virtual Reality and 3D User
  Interfaces (VR)}, pp. 159--166. IEEE, 2018.

\bibitem{guo2019mixed}
J.~Guo, D.~Weng, Z.~Zhang, H.~Jiang, Y.~Liu, Y.~Wang, and H.~B.-L. Duh.
\newblock Mixed reality office system based on maslow’s hierarchy of needs:
  Towards the long-term immersion in virtual environments.
\newblock In {\em 2019 IEEE International Symposium on Mixed and Augmented
  Reality (ISMAR)}, pp. 224--235. IEEE, 2019.

\bibitem{hasan2012acoord}
K.~Hasan, X.-D. Yang, A.~Bunt, and P.~Irani.
\newblock A-coord input: Coordinating auxiliary input streams for augmenting
  contextual pen-based interactions.
\newblock In {\em Proceedings of the SIGCHI Conference on Human Factors in
  Computing Systems}, CHI ’12, p. 805–814. Association for Computing
  Machinery, New York, NY, USA, 2012. doi: {{%
10\hspace{.1pt}\discretionary{.}{%
}{.}\hspace{.4pt}1145\discretionary{/}{%
}{/}2207676\hspace{.1pt}\discretionary{.}{%
}{.}\hspace{.4pt}2208519}}


\bibitem{Hermans2011}
F.~Hermans, M.~Pinzger, and A.~van Deursen.
\newblock Supporting professional spreadsheet users by generating leveled
  dataflow diagrams.
\newblock In {\em Proceedings of the 33rd International Conference on Software
  Engineering}, ICSE ’11, p. 451–460. Association for Computing Machinery,
  2011. doi: {{%
10\hspace{.1pt}\discretionary{.}{%
}{.}\hspace{.4pt}1145\discretionary{/}{%
}{/}1985793\hspace{.1pt}\discretionary{.}{%
}{.}\hspace{.4pt}1985855}}


\bibitem{hilliges2009interactions}
O.~Hilliges, S.~Izadi, A.~Wilson, S.~Hodges, A.~Garcia-Mendoza, and A.~Butz.
\newblock Interactions in the air: Adding further depth to interactive
  tabletops.
\newblock In {\em UIST '09 Proceedings of the 22nd annual ACM symposium on User
  interface software and technology}, pp. 139--148. ACM, October 2009.

\bibitem{hinckley2013motion}
K.~Hinckley, X.~A. Chen, and H.~Benko.
\newblock Motion and context sensing techniques for pen computing.
\newblock In {\em Proceedings of Graphics Interface 2013}, GI ’13, p.
  71–78. Canadian Information Processing Society, CAN, 2013.

\bibitem{hinckley2016pre-touch}
K.~Hinckley, S.~Heo, M.~Pahud, C.~Holz, H.~Benko, A.~Sellen, R.~Banks,
  K.~O'Hara, G.~Smyth, and B.~Buxton.
\newblock Pre-touch sensing for mobile interaction.
\newblock In {\em CHI '16 Proceedings of the 2016 CHI Conference on Human
  Factors in Computing Systems}, pp. 2869--2881. ACM, May 2016.

\bibitem{hinckley2014sensing}
K.~Hinckley, M.~Pahud, H.~Benko, P.~Irani, F.~Guimbreti\`{e}re, M.~Gavriliu,
  X.~A. Chen, F.~Matulic, W.~Buxton, and A.~Wilson.
\newblock Sensing techniques for tablet+stylus interaction.
\newblock In {\em Proceedings of the 27th Annual ACM Symposium on User
  Interface Software and Technology}, UIST ’14, p. 605–614. Association for
  Computing Machinery, New York, NY, USA, 2014. doi: {{%
10\hspace{.1pt}\discretionary{.}{%
}{.}\hspace{.4pt}1145\discretionary{/}{%
}{/}2642918\hspace{.1pt}\discretionary{.}{%
}{.}\hspace{.4pt}2647379}}


\bibitem{hinckley2010pen}
K.~Hinckley, K.~Yatani, M.~Pahud, N.~Coddington, J.~Rodenhouse, A.~Wilson,
  H.~Benko, and B.~Buxton.
\newblock Pen + touch = new tools.
\newblock In {\em UIST '10 Proceedings of the 23nd annual ACM symposium on User
  interface software and technology}, pp. 27--36. ACM, October 2010.

\bibitem{hinckley2007inkseine}
K.~Hinckley, S.~Zhao, R.~Sarin, P.~Baudisch, E.~Cutrell, M.~Shilman, and
  D.~Tan.
\newblock Inkseine: In situ search for active note taking.
\newblock In {\em CHI '07 Proceedings of the SIGCHI Conference on Human Factors
  in Computing Systems}, pp. 251--260. ACM, April 2007.

\bibitem{hwang2013magpen}
S.~Hwang, A.~Bianchi, M.~Ahn, and K.~Wohn.
\newblock Magpen: Magnetically driven pen interactions on and around
  conventional smartphones.
\newblock In {\em Proceedings of the 15th International Conference on
  Human-Computer Interaction with Mobile Devices and Services}, MobileHCI
  ’13, p. 412–415. Association for Computing Machinery, New York, NY, USA,
  2013. doi: {{%
10\hspace{.1pt}\discretionary{.}{%
}{.}\hspace{.4pt}1145\discretionary{/}{%
}{/}2493190\hspace{.1pt}\discretionary{.}{%
}{.}\hspace{.4pt}2493194}}


\bibitem{jannach2016model}
D.~Jannach and T.~Schmitz.
\newblock Model-based diagnosis of spreadsheet programs: a constraint-based
  debugging approach.
\newblock {\em Automated Software Engineering}, 23(1):105--144, 2016.

\bibitem{jeong2016ergonomic}
S.~Jeong, E.~S. Jung, and Y.~Im.
\newblock Ergonomic evaluation of interaction techniques and 3d menus for the
  practical design of 3d stereoscopic displays.
\newblock {\em International Journal of Industrial Ergonomics}, 53:205--218,
  2016.

\bibitem{jones2003user}
S.~P. Jones, A.~Blackwell, and M.~Burnett.
\newblock A user-centred approach to functions in excel.
\newblock In {\em Proceedings of the eighth ACM SIGPLAN international
  conference on Functional programming}, pp. 165--176, 2003.

\bibitem{kandogan2005a1}
E.~Kandogan, E.~Haber, R.~Barrett, A.~Cypher, P.~Maglio, and H.~Zhao.
\newblock A1: end-user programming for web-based system administration.
\newblock In {\em Proceedings of the 18th annual ACM symposium on User
  interface software and technology}, pp. 211--220, 2005.

\bibitem{knierim2018physical}
P.~Knierim, V.~Schwind, A.~M. Feit, F.~Nieuwenhuizen, and N.~Henze.
\newblock Physical keyboards in virtual reality: Analysis of typing performance
  and effects of avatar hands.
\newblock In {\em Proceedings of the 2018 CHI Conference on Human Factors in
  Computing Systems}, p. 345. ACM, 2018.

\bibitem{kobayashi1998enhanceddesk}
M.~Kobayashi and H.~Koike.
\newblock Enhanceddesk: integrating paper documents and digital documents.
\newblock In {\em Proceedings. 3rd Asia Pacific Computer Human Interaction
  (Cat. No. 98EX110)}, pp. 57--62. IEEE, 1998.

\bibitem{kohlhase2015context}
A.~Kohlhase, M.~Kohlhase, and A.~Guseva.
\newblock Context in spreadsheet comprehension.
\newblock In {\em SEMS@ ICSE}, pp. 21--27, 2015.

\bibitem{komerska2004study}
R.~Komerska and C.~Ware.
\newblock A study of haptic linear and pie menus in a 3d fish tank vr
  environment.
\newblock In {\em 12th International Symposium on Haptic Interfaces for Virtual
  Environment and Teleoperator Systems, 2004. HAPTICS'04. Proceedings.}, pp.
  224--231. IEEE, 2004.

\bibitem{kry2008handnavigator}
P.~G. Kry, A.~Pihuit, A.~Bernhardt, and M.-P. Cani.
\newblock Handnavigator: Hands-on interaction for desktop virtual reality.
\newblock In {\em Proceedings of the 2008 ACM symposium on Virtual reality
  software and technology}, pp. 53--60. ACM, 2008.

\bibitem{kulshreshth2014exploring}
A.~Kulshreshth and J.~J. LaViola~Jr.
\newblock Exploring the usefulness of finger-based 3d gesture menu selection.
\newblock In {\em Proceedings of the SIGCHI Conference on Human Factors in
  Computing Systems}, pp. 1093--1102, 2014.

\bibitem{Kurtenbach94userlearning}
G.~Kurtenbach and W.~Buxton.
\newblock User learning and performance with marking menus.
\newblock pp. 258--264. ACM Press, 1994.

\bibitem{li2020grip}
N.~Li, T.~Han, F.~Tian, Huang, S.~Jin, P.~Minhui, Irani, and J.~Alexander.
\newblock Get a grip: Evaluating grip gestures for vr input using a lightweight
  pen.
\newblock In {\em Proceedings of the 2020 CHI Conference on Human Factors in
  Computing Systems}. ACM, 2020.

\bibitem{li2019holodoc}
Z.~Li, M.~Annett, K.~Hinckley, K.~Singh, and D.~Wigdor.
\newblock Holodoc: Enabling mixed reality workspaces that harness physical and
  digital content.
\newblock In {\em Proceedings of the 2019 CHI Conference on Human Factors in
  Computing Systems}, p. 687. ACM, 2019.

\bibitem{lim2019evaluation}
Z.~H. Lim and P.~O. Kristensson.
\newblock An evaluation of discrete and continuous mid-air loop and marking
  menu selection in optical see-through hmds.
\newblock In {\em Proceedings of the 21st International Conference on
  Human-Computer Interaction with Mobile Devices and Services}, pp. 1--10,
  2019.

\bibitem{liu2012flexaura}
S.~Liu and F.~Guimbreti\`{e}re.
\newblock Flexaura: A flexible near-surface range sensor.
\newblock In {\em Proceedings of the 25th Annual ACM Symposium on User
  Interface Software and Technology}, UIST ’12, p. 327–330. Association for
  Computing Machinery, New York, NY, USA, 2012. doi: {{%
10\hspace{.1pt}\discretionary{.}{%
}{.}\hspace{.4pt}1145\discretionary{/}{%
}{/}2380116\hspace{.1pt}\discretionary{.}{%
}{.}\hspace{.4pt}2380158}}


\bibitem{mack2018characterizing}
K.~Mack, J.~Lee, K.~Chang, K.~Karahalios, and A.~Parameswaran.
\newblock Characterizing scalability issues in spreadsheet software using
  online forums.
\newblock In {\em Extended Abstracts of the 2018 CHI Conference on Human
  Factors in Computing Systems}, pp. 1--9, 2018.

\bibitem{mackay2002interactionTechnique}
W.~Mackay.
\newblock Which interaction technique works when?: floating palettes, marking
  menus and toolglasses support different task strategies.
\newblock pp. 203--208, 01 2002. doi: {{%
10\hspace{.1pt}\discretionary{.}{%
}{.}\hspace{.4pt}1145\discretionary{/}{%
}{/}1556262\hspace{.1pt}\discretionary{.}{%
}{.}\hspace{.4pt}1556294}}


\bibitem{marquardt2011continuous}
N.~Marquardt, R.~Jota, S.~Greenberg, and J.~A. Jorge.
\newblock The continuous interaction space: interaction techniques unifying
  touch and gesture on and above a digital surface.
\newblock In {\em IFIP Conference on Human-Computer Interaction}, pp. 461--476.
  Springer, 2011.

\bibitem{matulic2020pensight}
F.~Matulic, R.~Arakawa, B.~Vogel, and D.~Vogel.
\newblock Pensight: Enhanced interaction with a pen-top camera.
\newblock In {\em Proceedings of the 2020 CHI Conference on Human Factors in
  Computing Systems}. ACM, 2020.

\bibitem{matulic2013pen}
F.~Matulic and M.~C. Norrie.
\newblock Pen and touch gestural environment for document editing on
  interactive tabletops.
\newblock In {\em Proceedings of the 2013 ACM international conference on
  Interactive tabletops and surfaces}, pp. 41--50, 2013.

\bibitem{mcgill2015dose}
M.~McGill, D.~Boland, R.~Murray-Smith, and S.~Brewster.
\newblock A dose of reality: Overcoming usability challenges in vr head-mounted
  displays.
\newblock In {\em Proceedings of the 33rd Annual ACM Conference on Human
  Factors in Computing Systems}, pp. 2143--2152. ACM, 2015.

\bibitem{mcgill2019challenges}
M.~McGill, J.~Williamson, A.~Ng, F.~Pollick, and S.~Brewster.
\newblock Challenges in passenger use of mixed reality headsets in cars and
  other transportation.
\newblock {\em Virtual Reality}, pp. 1--21, 2019.

\bibitem{miller2016gradual}
G.~Miller and F.~Hermans.
\newblock Gradual structuring in the spreadsheet paradigm.
\newblock In {\em 2016 IEEE Symposium on Visual Languages and Human-Centric
  Computing (VL/HCC)}, pp. 240--241. IEEE, 2016.

\bibitem{monteiro2019comparison}
P.~Monteiro, H.~Coelho, G.~Gon{\c{c}}alves, M.~Melo, and M.~Bessa.
\newblock Comparison of radial and panel menus in virtual reality.
\newblock {\em IEEE Access}, 7:116370--116379, 2019.

\bibitem{panko2016we}
R.~Panko.
\newblock What we don't know about spreadsheet errors today: The facts, why we
  don't believe them, and what we need to do.
\newblock {\em arXiv preprint arXiv:1602.02601}, 2016.

\bibitem{pfeuffer2017thumb}
K.~Pfeuffer, K.~Hinckley, M.~Pahud, and B.~Buxton.
\newblock Thumb + pen interaction on tablets.
\newblock In {\em Proceedings of the 2017 CHI Conference on Human Factors in
  Computing Systems}, CHI ’17, p. 3254–3266. Association for Computing
  Machinery, New York, NY, USA, 2017. doi: {{%
10\hspace{.1pt}\discretionary{.}{%
}{.}\hspace{.4pt}1145\discretionary{/}{%
}{/}3025453\hspace{.1pt}\discretionary{.}{%
}{.}\hspace{.4pt}3025567}}


\bibitem{pham2019pen}
D.-M. Pham and W.~Stuerzlinger.
\newblock Is the pen mightier than the controller? a comparison of input
  devices for selection in virtual and augmented reality.
\newblock In {\em 25th ACM Symposium on Virtual Reality Software and
  Technology}, pp. 1--11, 2019.

\bibitem{pinhanez2001everywhere}
C.~Pinhanez.
\newblock The everywhere displays projector: A device to create ubiquitous
  graphical interfaces.
\newblock In {\em International conference on ubiquitous computing}, pp.
  315--331. Springer, 2001.

\bibitem{pollock2019countmarks}
J.~R.~F. Pollock.
\newblock {\em CountMarks: Multi-Finger Marking Menus for Mobile Interaction
  with Head-Mounted Displays}.
\newblock PhD thesis, Carleton University, 2019.

\bibitem{raymond1986survey}
D.~R. Raymond.
\newblock {\em A survey of research in computer-based menus}.
\newblock Citeseer, 1986.

\bibitem{rekimoto1999augmented}
J.~Rekimoto and M.~Saitoh.
\newblock Augmented surfaces: a spatially continuous work space for hybrid
  computing environments.
\newblock In {\em Proceedings of the SIGCHI conference on Human Factors in
  Computing Systems}, pp. 378--385. ACM, 1999.

\bibitem{ruvimova2020transport}
A.~Ruvimova, J.~Kim, T.~Fritz, M.~Hancock, and D.~C. Shepherd.
\newblock "transport me away": Fostering flow in open offices through virtual
  reality.
\newblock In {\em Proceedings of the 2020 CHI Conference on Human Factors in
  Computing Systems}. ACM, 2020.

\bibitem{schneider2019reconviguration}
D.~Schneider, A.~Otte, T.~Gesslein, P.~Gagel, B.~Kuth, M.~S. Damlakhi,
  O.~Dietz, E.~Ofek, M.~Pahud, P.~O. Kristensson, et~al.
\newblock Reconviguration: Reconfiguring physical keyboards in virtual reality.
\newblock {\em IEEE transactions on visualization and computer graphics}, 2019.

\bibitem{schneider2020accuracy}
D.~Schneider, A.~Otte, A.~S. Kublin, A.~Martschenko, P.~O. Kristensson,
  E.~Ofek, M.~Pahud, and J.~Grubert.
\newblock Accuracy of commodity finger tracking systems for virtual reality
  head-mounted displays.
\newblock In {\em 2020 IEEE Conference on Virtual Reality and 3D User
  Interfaces Abstracts and Workshops (VRW)}, pp. 805--806. IEEE, 2020.

\bibitem{Shiozawa1999}
H.~Shiozawa, K.-i. Okada, and Y.~Matsushita.
\newblock 3d interactive visualization for inter-cell dependencies of
  spreadsheets.
\newblock In {\em Proceedings of the 1999 IEEE Symposium on Information
  Visualization}, INFOVIS ’99, p.~79. IEEE Computer Society, 1999.

\bibitem{smith2017spreadsheet}
J.~Smith, J.~A. Middleton, and N.~A. Kraft.
\newblock Spreadsheet practices and challenges in a large multinational
  conglomerate.
\newblock In {\em 2017 IEEE Symposium on Visual Languages and Human-Centric
  Computing (VL/HCC)}, pp. 155--163. IEEE, 2017.

\bibitem{hyong2011grips}
H.~Song, H.~Benko, F.~Guimbretiere, S.~Izadi, X.~Cao, and K.~Hinckley.
\newblock Grips and gestures on a multi-touch pen.
\newblock In {\em Proceedings of the SIGCHI Conference on Human Factors in
  Computing Systems}, CHI ’11, p. 1323–1332. Association for Computing
  Machinery, New York, NY, USA, 2011. doi: {{%
10\hspace{.1pt}\discretionary{.}{%
}{.}\hspace{.4pt}1145\discretionary{/}{%
}{/}1978942\hspace{.1pt}\discretionary{.}{%
}{.}\hspace{.4pt}1979138}}


\bibitem{speicher2018VRselection}
M.~Speicher, A.~M. Feit, P.~Ziegler, and A.~Kr\"{u}ger.
\newblock Selection-based text entry in virtual reality.
\newblock In {\em Proceedings of the 2018 CHI Conference on Human Factors in
  Computing Systems}, CHI ’18, p. 1–13. Association for Computing
  Machinery, New York, NY, USA, 2018. doi: {{%
10\hspace{.1pt}\discretionary{.}{%
}{.}\hspace{.4pt}1145\discretionary{/}{%
}{/}3173574\hspace{.1pt}\discretionary{.}{%
}{.}\hspace{.4pt}3174221}}


\bibitem{surale2019tabletinvr}
H.~B. Surale, A.~Gupta, M.~Hancock, and D.~Vogel.
\newblock Tabletinvr: Exploring the design space for using a multi-touch tablet
  in virtual reality.
\newblock In {\em Proceedings of the 2019 CHI Conference on Human Factors in
  Computing Systems}, p.~13. ACM, 2019.

\bibitem{suzuki2009interaction}
Y.~Suzuki, K.~Misue, and J.~Tanaka.
\newblock Interaction technique for a pen-based interface using finger motions.
\newblock In {\em International Conference on Human-Computer Interaction}, pp.
  503--512. Springer, 2009.

\bibitem{syrdal2008video}
D.~S. Syrdal, N.~Otero, and K.~Dautenhahn.
\newblock Video prototyping in human-robot interaction: Results from a
  qualitative study.
\newblock In {\em Proceedings of the 15th European conference on Cognitive
  ergonomics: the ergonomics of cool interaction}, pp. 1--8, 2008.

\bibitem{szalavari1997personal}
Z.~Szalav{\'a}ri and M.~Gervautz.
\newblock The personal interaction panel--a two-handed interface for augmented
  reality.
\newblock In {\em Computer graphics forum}, vol.~16, pp. C335--C346. Wiley
  Online Library, 1997.

\bibitem{szalavari1997using}
Z.~Szalav{\'a}ri and M.~Gervautz.
\newblock Using the personal interaction panel for 3d interaction.
\newblock In {\em proceedings of the Conference on Latest Results in
  Information Technology}, p.~36. Citeseer, 1997.

\bibitem{teyssier2017versapen}
M.~Teyssier, G.~Bailly, and E.~Lecolinet.
\newblock Versapen: An adaptable, modular and multimodal i/o pen.
\newblock In {\em Proceedings of the 2017 CHI Conference Extended Abstracts on
  Human Factors in Computing Systems}, CHI EA ’17, p. 2155–2163.
  Association for Computing Machinery, New York, NY, USA, 2017. doi: {{%
10\hspace{.1pt}\discretionary{.}{%
}{.}\hspace{.4pt}1145\discretionary{/}{%
}{/}3027063\hspace{.1pt}\discretionary{.}{%
}{.}\hspace{.4pt}3053159}}


\bibitem{tian2008tilt}
F.~Tian, L.~Xu, H.~Wang, X.~Zhang, Y.~Liu, V.~Setlur, and G.~Dai.
\newblock Tilt menu: Using the 3d orientation information of pen devices to
  extend the selection capability of pen-based user interfaces.
\newblock In {\em Proceedings of the SIGCHI Conference on Human Factors in
  Computing Systems}, CHI ’08, p. 1371–1380. Association for Computing
  Machinery, New York, NY, USA, 2008. doi: {{%
10\hspace{.1pt}\discretionary{.}{%
}{.}\hspace{.4pt}1145\discretionary{/}{%
}{/}1357054\hspace{.1pt}\discretionary{.}{%
}{.}\hspace{.4pt}1357269}}


\bibitem{tu2019crossing}
H.~Tu, S.~Huang, J.~Yuan, X.~Ren, and F.~Tian.
\newblock Crossing-based selection with virtual reality head-mounted displays.
\newblock In {\em Proceedings of the 2019 CHI Conference on Human Factors in
  Computing Systems}, pp. 1--14, 2019.

\bibitem{unger2012project}
R.~Unger and C.~Chandler.
\newblock {\em A Project Guide to UX Design: For user experience designers in
  the field or in the making}.
\newblock New Riders, 2012.

\bibitem{wagner2018virtualdesk}
J.~A. Wagner~Filho, C.~M. D.~S. Freitas, and L.~Nedel.
\newblock Virtualdesk: a comfortable and efficient immersive information
  visualization approach.
\newblock In {\em Computer Graphics Forum}, vol.~37, pp. 415--426. Wiley Online
  Library, 2018.

\bibitem{webb2016wearables}
A.~Webb, M.~Pahud, K.~Hickley, and B.~Buxton.
\newblock Wearables as context for guiard-abiding bimanual touch.
\newblock In {\em UIST '16 Proceedings of the 29th Annual Symposium on User
  Interface Software and Technology}, pp. 287--300. ACM, October 2016.

\bibitem{wellner1994interacting}
P.~D. Wellner.
\newblock Interacting with paper on the digitaldesk.
\newblock Technical report, University of Cambridge, Computer Laboratory, 1994.

\bibitem{white2009interaction}
S.~White, D.~Feng, and S.~Feiner.
\newblock Interaction and presentation techniques for shake menus in tangible
  augmented reality.
\newblock In {\em 2009 8th IEEE International Symposium on Mixed and Augmented
  Reality}, pp. 39--48. IEEE, 2009.

\bibitem{xia2017writlarge}
H.~Xia, K.~Hinckley, M.~Pahud, X.~Tu, and B.~Buxton.
\newblock Writlarge: Ink unleashed by unified scope, action, and zoom.
\newblock In {\em Proceedings of the 2017 CHI Conference on Human Factors in
  Computing Systems (CHI '17)}. ACM, May 2017.
\newblock Honorable Mention.

\bibitem{zhang2019sensing}
Y.~Zhang, M.~Pahud, C.~Holz, H.~Xia, G.~Laput, M.~McGuffin, X.~Tu,
  A.~Mittereder, F.~Su, B.~Buxton, and K.~Hinckley.
\newblock Sensing posture-aware pen+touch interaction on tablets.
\newblock In {\em CHI 2019 Conference on Human Factors in Computing Systems}.
  ACM, May 2019.

\bibitem{zielasko2019menus}
D.~Zielasko, M.~Kr{\"u}ger, B.~Weyers, and T.~W. Kuhlen.
\newblock Menus on the desk? system control in deskvr.
\newblock In {\em 2019 IEEE Conference on Virtual Reality and 3D User
  Interfaces (VR)}, pp. 1287--1288. IEEE, 2019.

\bibitem{zielasko2019passive}
D.~Zielasko, M.~Kr{\"u}ger, B.~Weyers, and T.~W. Kuhlen.
\newblock Passive haptic menus for desk-based and hmd-projected virtual
  reality.
\newblock In {\em 2019 IEEE 5th Workshop on Everyday Virtual Reality (WEVR)},
  pp. 1--6. IEEE, 2019.

\end{thebibliography}
\end{document}